\documentclass[a4paper,12pt]{article}
\pdfoutput=1 

\usepackage{jheppub} 

\usepackage{tabularx,subcaption,fancyhdr,mathtools}
\usepackage[makeroom]{cancel}
\usepackage{bm}
\usepackage{csquotes}
\usepackage[nottoc,notlot,notlof]{tocbibind}
\usepackage{float}
\usepackage{relsize}
\usepackage{hyperref}
\usepackage{color,latexsym, array,multirow,url,verbatim, enumerate,cancel}
\DeclareUnicodeCharacter{2212}{-}
\hypersetup{colorlinks=true,linkcolor=blue,citecolor=magenta,filecolor=magenta,
	urlcolor=cyan}
\DeclareUnicodeCharacter{2212}{-}
\usepackage{setspace}
\usepackage{colortbl}
\usepackage{cancel}
\usepackage{xcolor}
\usepackage{multicol}
\usepackage{graphicx}
\usepackage[normalem]{ulem}
\usepackage{longtable}

\newcommand{\beq}{\begin{equation}}
\newcommand{\eeq}{\end{equation}}
\def\bea{\begin{eqnarray}}
\def\eea{\end{eqnarray}}

\title{Controls on the deformation and reorientation of mafic enclaves during magma mingling: A computational study}

\author[a]{Pragati Sonali,} 
\author[b]{Arghya Choudhury,}

\author[a,*]{and Urmi Dutta%
\begingroup
\footnotetext[1]{urmidutta.geology@gmail.com (corresponding author)}%
\endgroup}

\affiliation[a]{Department of Geology, Patna University, Bihar - 800005, India}
\affiliation[b]{Department of Physics, Indian Institute of Technology Patna, Bihar - 801106, India}

\abstract{Mafic microgranular enclaves in granitic plutons record magma mingling, and their shapes are often interpreted as a record of the flow that deformed them. We investigate how a dense, low-viscosity droplet deforms in a lighter, much more viscous host under shear and gravity, using 96 two-dimensional Level-Set simulations. The viscosity ratio ($\mu_r$) spans $10^{-6}$ - $10^{-3}$, the capillary number ($\mathrm{Ca}$) ranges from $10^{6}$ to $8\times10^{9}$, the Bond number ($\mathrm{Bo}$) takes values of $3.5\times10^{6}$ and $9.8\times10^{6}$, and the Reynolds number ($\mathrm{Re}$) remains below $10^{-2}$, with a fixed density contrast ($\Delta\rho$) of $400~\mathrm{kg\,m^{-3}}$. The vertical confinement ratio ($2a/H$) is either 0.12 or 0.20, and the lateral confinement ratio ($2a/L$) is varied between 0.075 and 0.125. Two broad regimes identified- sinking and non-sinking which spread over six outcomes like sinking, edge, mid and multiple break-up, rotation, and retraction. The ratio $\mathrm{Ca/Bo}$ compares the imposed shear  with the gravity induced settling and does not depend on interfacial tension. It separates sinking from shear-dominated behaviour near unity, although the transition spans a range of $\mathrm{Ca/Bo}$ values that shifts with confinement. In the shear-dominated regime, a lower viscosity ratio promotes stretching. Multiple pinch-off occurs mainly at viscosity ratios of $10^{-6}$ - $10^{-5}$, rotation occurs only at the higher vertical confinement, and retraction occurs only at the lower vertical confinement. The strain at maximum stretch averages 1.1 - 1.4 and increases with $\mathrm{Ca/Bo}$, with a stronger increase at higher confinement. The orientation of the major axis decreases by approximately $10^\circ$ per unit strain and is lowest for rotating droplets. Because the capillary number exceeds conventional critical values for breakup for droplets, breakup is interpreted in terms of accumulated deformation rather than capillary stability alone. Enclave shape therefore reflects the flow regime and is not a direct measure of finite strain.}


\begin{document} 

\maketitle

\section{Introduction}
Granitic plutons are one of the most common intrusive features on Earth and their widespread occurrences record the evolution of the continental crust. Mafic microgranular enclaves  in granitic plutons are commonly interpreted as fragments of injected mafic magma that became dispersed within a felsic host during magma mingling \citep{barbarin1992genesis, barbarin2005mafic}. Depending on the physical conditions prevailing during emplacement, enclaves may remain nearly spherical, become progressively elongated, fragment into smaller bodies, or develop highly irregular morphologies. A review of microgranitoid enclave shapes and orientations concluded that these features need careful interpretation and cannot simply be treated as strain markers \cite{paterson2004significance}, and multi-pronged enclaves in particular have been used to argue that several fabrics can be recorded in the same pluton \cite{paterson2003quadruple}. The deformed enclave shapes, in field, range from rounded and ellipsoidal bodies to strongly elongated, folded or fragmented forms, including chains of small fragments and multi-pronged bodies. A systematic study of which flow conditions produce which enclave shapes is still pending. 
A mafic enclave in a felsic host is a drop of one viscous fluid in another, so the classical theory of drop deformation applies \cite{taylor1932viscosity, taylor1934formation, rallison1984deformation,stone1994dynamics,guido2011shear}.

For Newtonian fluids in simple shear, the response is controlled by a small set of dimensionless groups: the capillary number $\mathrm{Ca}$, which compares viscous and interfacial stresses, the viscosity ratio $\mu_r$ between drop and host, the Reynolds number $\mathrm{Re}$, and, when walls are close, the confinement ratio $2a/H$ of drop diameter to wall separation. At small $\mathrm{Ca}$ a circular drop deforms into an ellipse inclined at roughly $45^{\circ}$ to the flow \cite{taylor1934formation}. With increasing $\mathrm{Ca}$ it elongates and turns towards the flow direction until it either reaches a steady shape or breaks up.
Confinement modifies this picture. Lattice-Boltzmann simulations of confined shear flow find that stronger confinement increases deformation and turns the drop further towards the flow direction, and the same trend has been reported in experiments and in other simulations \citep{vananroye2007effect,hua2013effect}. Volume-of-Fluid, phase-field, lattice-Boltzmann and Level-Set methods can resolve strongly deformed complex interfaces \citep{dutta2014ascent,komrakova2014lattice,dutta2016role,amani2019dns,wang2023dynamics,dutta2024solitary}. \citep{amani2019dns} used a Level-Set method to map steady and break-up regions across $\mathrm{Ca}$, $\mu_r$, $\mathrm{Re}$ and confinement ratio, and \citep{wang2023dynamics} used a multiphase SPH method to show that deformation grows with both $\mathrm{Ca}$ and $\mathrm{Re}$ and that the critical $\mathrm{Ca}$ for break-up depends on confinement.

\begin{figure}[!htb]
\begin{center}
\includegraphics[angle=0, width=0.98\textwidth]{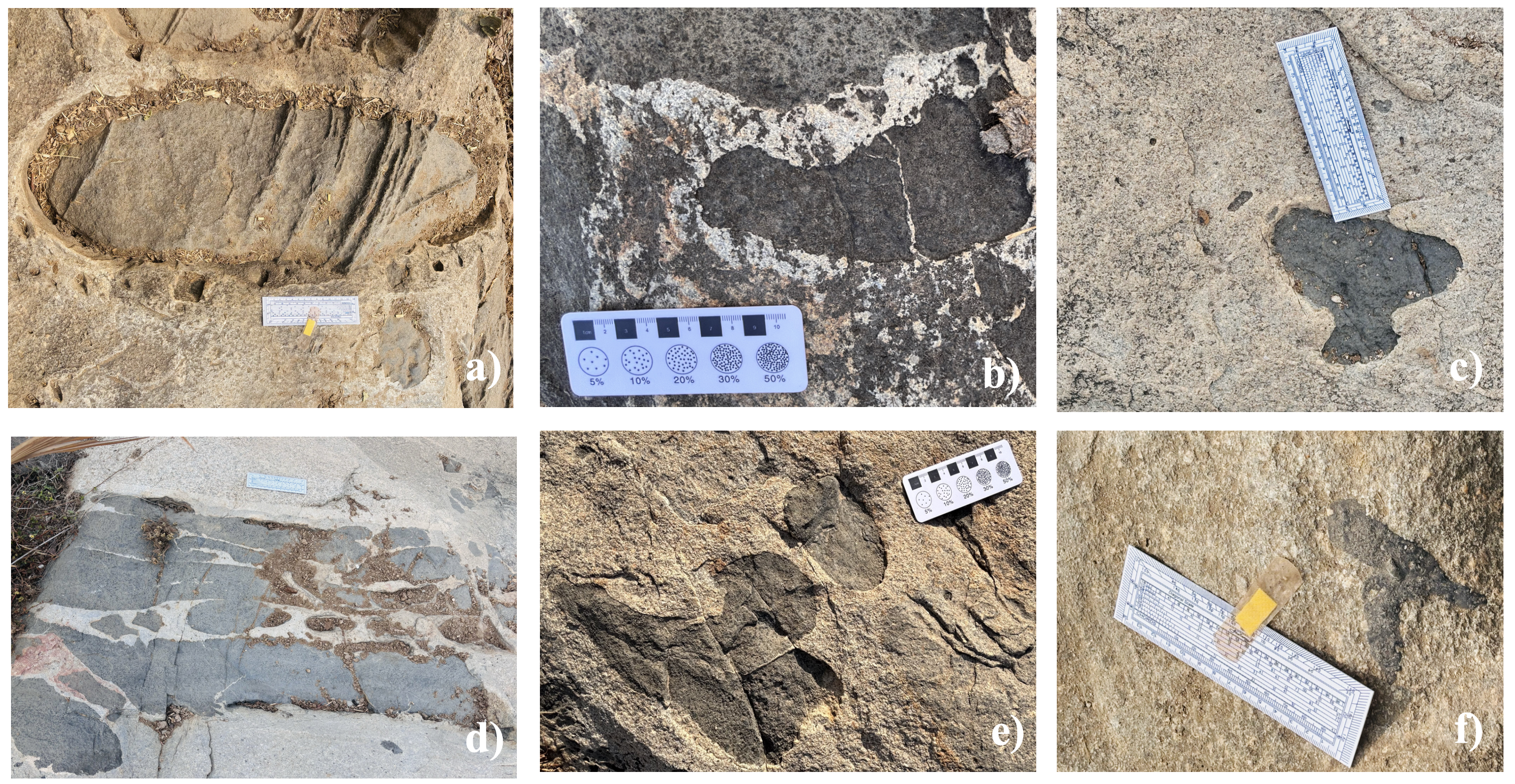}
\caption{Different shapes of deformed enclaves found in Nimchak Pluton:  a) elliptical; b) \& e) curved; c) multi-pronged; d) flattened; f) folded.}
\label{fig:fig1}
\end{center}
\end{figure}

Most of these works concern viscosity ratios of order one and capillary numbers close to the critical value for break-up. The viscosity ratio nevertheless matters a great deal. When the host is no more viscous than the drop, deformation is sluggish and a higher $\mathrm{Ca}$ is needed to reach a given shape. When the drop is much less viscous than the host, vigorous internal circulation allows deformation at modest $\mathrm{Ca}$ and leads to pinch-off at the tips \citep{stone1989relaxation,guido2011shear}. In highly viscous magmatic system with negligible inertia and surface tension, extreme values of $\mathrm{Ca}$  and $\mathrm{Re}$ are expected. Break-ups and other deformation of viscous drop, at such high parameter range, have hardly been explored. Similarly, negetive buoyancy effect on droplets has also received very little attention. In the engineering and biological flows that motivate most droplet studies, settling is too slow to compete with shear, and the body-force term is usually omitted from the governing equations. Bond-number effects have been examined for sheared drops attached to a wall \citep{dimitrakopoulos2007gravitational}, but we are not aware of a systematic study of how settling and shear compete for a free drop over a controlled parameter space. In magmatic systems both processes act over long times, and their balance is measured by $\mathrm{Ca}/\mathrm{Bo} = \mu_h U/(\Delta\rho\, g\, a^{2})$, where $\mu_h$ is the host viscosity, $U$ the characteristic shear velocity, $\Delta\rho$ the density contrast and $a$ the drop radius. Up to a factor of order unity, $\mathrm{Ca}/\mathrm{Bo}$ is the ratio of the imposed shear to the gravity induced settling  of the drop, and because interfacial tension cancels, it does not depend on $\sigma$.
Natural magmas lie in exactly this regime. Felsic granitic melts have viscosities of $10^{5}$ - $10^{9}$~Pa\,s \citep{petford2003rheology}, and the mafic magmas injected into them are several orders of magnitude less viscous, of order $10$ - $10^{3}$~Pa\,s \citep{dingwell1996volcanic,giordano2008viscosity}. The  drop-to-host viscosity ratio can therefore reach $10^{-3}$ - $10^{-6}$. Interfacial tension between silicate melts is also recorded to be very low \citep{edwards2020magma,toramaru2023new}. Together with enclave sizes of centimetres to metres and these viscosities, this gives $\mathrm{Ca}\gg 1$, so interfacial tension offers essentially no resistance to deformation once flow is imposed, and the Reynolds number is far below unity, so inertia is negligible and the flow is Stokes-like \citep{glazner2014magmatic}. Mafic microgranular enclaves record the mingling of mafic and felsic magmas in plutons, as in the Nimchak pluton, India (Fig.~\ref{fig:fig1}). The two magmas differ in density as well as viscosity, and the interaction lasts long enough for buoyancy to act alongside shear resulting different deformed enclave shape.
Several lines of work have addressed enclave deformation. \citep{williams1994microgranitic} applied drop-deformation theory to deformation of microgranitic enclaves under the assumption of simple shear, \citep{rust2003determining} used the shapes and orientations of bubbles in obsidian to infer shear rate and flow type through the capillary number, and \citep{piazolo2002experimental} performed shear-rig experiments to get insight into different shape evolution of viscous inclusion. Analogue experiments on injected mafic dykes show that their break-up depends strongly on the rheological contrast between the magmas \citep{hodge2012experimental,hodge2012magmatic}, and \citep{burgisser2020numerical} simulated enclave deformation numerically and fitted a first-order relationship between shear and deformation. To our knowledge, none of these studies resolves how buoyancy and shear compete across a controlled parameter space at very high capillary number. Our earlier multiphase models using Volume-of-Fluid simulations of a less viscous fluid injected into a more viscous host showed that increasing the viscosity ratio changes the ascent from continuous to pulsating \citep{dutta2014ascent}. Here we extend such parametric study to the deformation of a dense droplet under shear and buoyancy.

We develop a two-dimensional Level-Set model of a dense, low-viscosity droplet embedded in a lighter, more viscous host between walls that move in opposite directions, and we run 96 simulations. The simulations span viscosity ratios $\mu_r = 10^{-6}$ - $10^{-3}$, capillary numbers $\mathrm{Ca} = 10^{6}$ - $8\times10^{9}$, Bond numbers $\mathrm{Bo} = 3.53\times10^{6}$ and $9.81\times10^{6}$ and Reynolds numbers $\mathrm{Re} = 10^{-6}$ - $10^{-2}$ , with a fixed density contrast $\Delta\rho = 400$~kg\,m$^{-3}$, vertical confinement ratios $2a/H = 0.12$ and $0.20$ and domain aspect ratios $L^{*} = 1.6$, $2$ and $3$. We separate a sinking regime, in which buoyancy dominates-the drop descends and gets flattened near the base, from a non-sinking regime, in which the drop stretches and then breaks up, rotates, or retracts. We show that the sinking transition is organised by $\mathrm{Ca}/\mathrm{Bo}$ and that confinement shifts it, and we relate the remaining outcomes to the viscosity ratio and confinement. We also quantify the orientation of the droplet and the strain at maximum stretch, which bear directly on whether enclave shape can serve as a strain marker.

Section~\ref{sec:method} describes the model and the mesh convergence. Section~\ref{sec:results} presents the deformation regimes, the regime map and the sinking transition, and the orientation and strain at maximum stretch. Section~\ref{sec:discussions} discusses the mechanisms and the implications for natural enclaves, and Section~\ref{sec:conclusions} gives the conclusions.



\section{Numerical Method}
\label{sec:method}


\providecommand{\toprule}{\hline}
\providecommand{\midrule}{\hline}
\providecommand{\bottomrule}{\hline}

\subsection{Model description}
\label{sec:model}

\begin{figure}[t]
  \centering
   \includegraphics[width=0.8\textwidth]{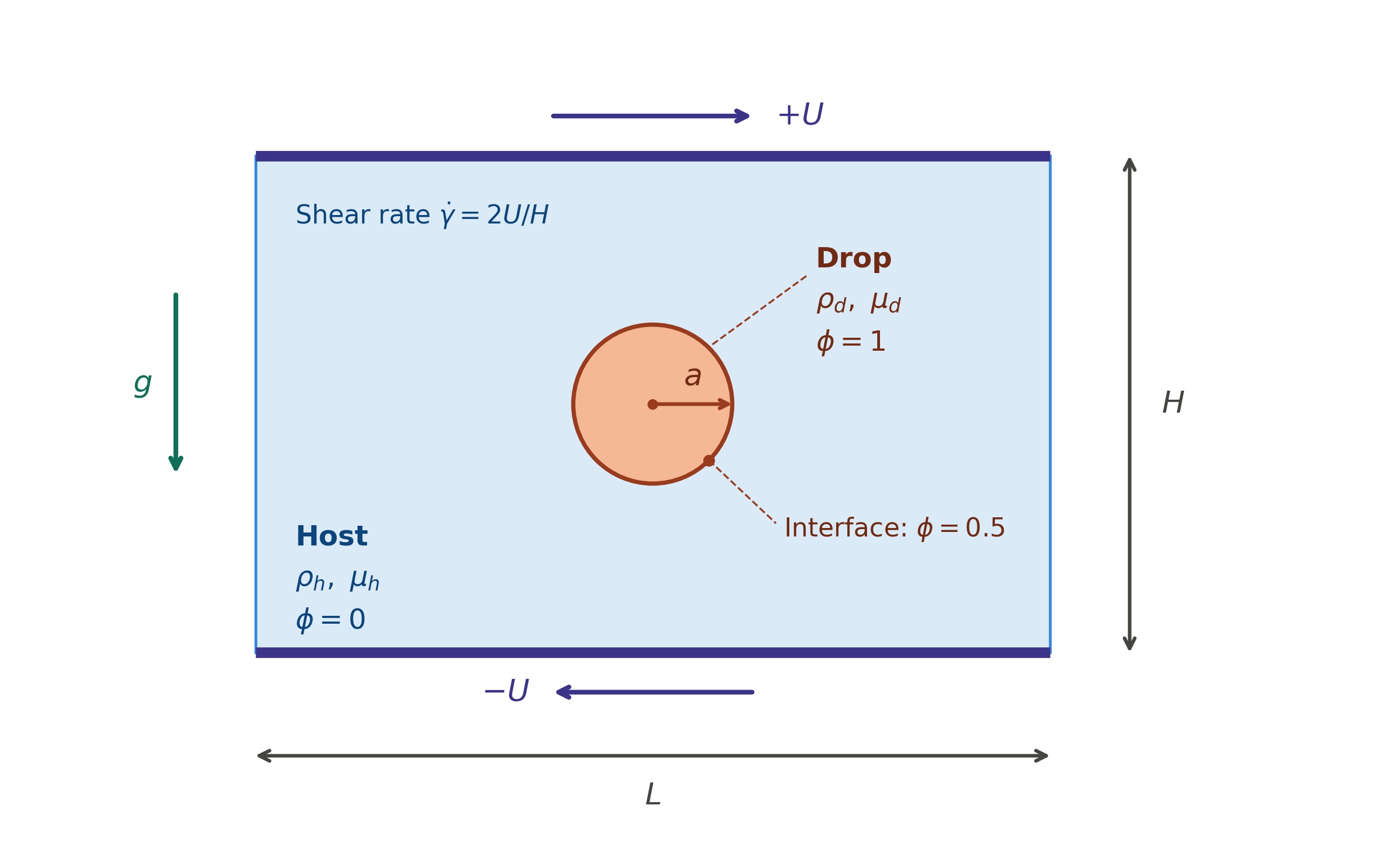}
    \caption{Schematic of the model setup (not to scale). A circular drop of radius $a$, density $\rho_d$ and viscosity $\mu_d$ is placed at the centre of a host of density $\rho_h$ and viscosity $\mu_h$ in a domain of length $L$ and height $H$. The level-set variable $\phi$ is 1 in the drop and 0 in the host, and the interface is the $\phi=0.5$ contour. The top and bottom walls move in opposite directions with speeds $\pm U$, imposing a shear of rate $\dot\gamma=2U/H$. Gravity acts in the negative $y$-direction.}
  \label{fig:setup}
\end{figure}

We consider a two-dimensional rectangular box of length $L$ and height $H$, with Cartesian coordinates $x\in[0,L]$ and $y\in[0,H]$ (Fig.~\ref{fig:setup}). A dense, low-viscosity circular drop (representing the mafic enclave) of radius $a$, density $\rho_d$ and viscosity $\mu_d$, is embedded in a lighter and much more viscous host (representing the felsic magma) of density $\rho_h$ and viscosity $\mu_h$. Gravity acts in the negative $y$-direction, $\mathbf{g}=-g\,\hat{\mathbf{y}}$, with $g=9.81~\mathrm{m\,s^{-2}}$. The top ($y=H$) and bottom ($y=0$) walls move in opposite directions with speeds $\pm U$, which imposes a nominal shear rate
$  \dot{\gamma} = \frac{2U}{H}$ on the host. No-slip conditions are applied at the moving walls, $\mathbf{u}(x,H)=(+U,0)$ and $\mathbf{u}(x,0)=(-U,0)$, whereas the left and right boundaries are free slip walls. All the walls thus have zero normal flux. The drop is initially at rest at the centre of the box, $(x_0,y_0)=(L/2,\,H/2)$. 
It is then subjected to the imposed shear flow and to downward gravitational motion arising from the density contrast \(\Delta\rho=\rho_d-\rho_h>0\).
The drop viscosity is held fixed at $\mu_d=10^3$~Pa\,s while the host viscosity is varied between $10^6$ and $10^9$~Pa\,s.  (Table~\ref{tab:parameters}).

\begin{table*}[htbp]
\centering
\footnotesize
\setlength{\tabcolsep}{5pt}
\renewcommand{\arraystretch}{1.2}
\begin{tabular}{|l|p{5.5cm}|p{3.2cm}|l|}
\hline
\rowcolor{gray!20}
\textbf{Parameter} & \textbf{Description} & \textbf{Range} & \textbf{Reference} \\ \hline
$\rho_d$ & Density of the magma drop & $2700~\text{kg m}^{-3}$ & \cite{petford2003rheology} \\ \hline
$\rho_h$ & Density of the host & $2300~\text{kg m}^{-3}$ & \cite{petford2003rheology} \\ \hline
$\Delta\rho$ & Density difference between the magma drop and the host & $400~\text{kg m}^{-3}$ &  \\ \hline
$\mu_d$ & Viscosity of the magma drop & $1000~\text{Pa}\cdot\text{s}$ & \cite{petford2003rheology} \\ \hline
$\mu_h$ & Viscosity of the host & $10^6 - 10^9~\text{Pa}\cdot\text{s}$ & \cite{petford2003rheology} \\ \hline
$\mu_r$ & Viscosity ratio of the host and the magma drop & $10^{-3}-10^{-6}$ &  \\ \hline
$\sigma$ & Interfacial tension of the silicic magma & $1.00 \times 10^{-2}$ & \cite{veksler2010interfacial} \\ \hline
\end{tabular}
\caption{Parameters, their physical ranges, and references used in the numerical model.}
\label{tab:parameters}
\end{table*}


\subsection{Governing equations}
\label{sec:governing}

The two magmas are treated as immiscible, incompressible Newtonian fluids with constant properties, and the flow is assumed to be isothermal. Following the one-fluid approach, a single set of equations is solved over the whole domain, with the fluid properties varying smoothly across a thin interfacial layer through the level-set (phase) variable $\phi(\mathbf{x},t)$, defined as $\phi=1$ in the drop and $\phi=0$ in the host. The flow is governed by the continuity and Navier-Stokes equations as follows:
\begin{equation}
  \nabla\cdot\mathbf{u} = 0,
  \label{eq:continuity}
\end{equation}
\begin{equation}
  \rho\left[\frac{\partial\mathbf{u}}{\partial t}
  + (\mathbf{u}\cdot\nabla)\mathbf{u}\right]
  = -\nabla p
  + \nabla\cdot\left[\mu\left(\nabla\mathbf{u}
  + \nabla\mathbf{u}^{T}\right)\right]
  + \rho\,\mathbf{g} + \mathbf{F}_{st},
  \label{eq:momentum}
\end{equation}
where $\mathbf{u}=(u,v)$ is the velocity, $p$ the pressure, $\mathbf{g}$ the gravitational acceleration and $\mathbf{F}_{st}$ the interfacial-tension force per unit volume. 

Following Brackbill et al.~\cite{brackbill1992}, the interfacial tension
enters as a body force acting in the interface region and expressed as:  
$  \mathbf{F}_{st} = \sigma\,\kappa\,\delta\,\mathbf{n}$, 
where $\sigma$ is the surface tension coefficient, $\kappa$ is the local curvature of the interface, $\delta$ is a regularised Dirac delta function that localises the surface tension force near the interface, and $\mathbf{n}$ is the unit normal vector to the interface. 
Surface tension is expected to have a limited influence in the present parameter range (capillary number  \(\mathrm{Ca}\gg1\)), but its contribution is retained in the governing equations.
The density and viscosity are interpolated across the interface as
\begin{equation}
  \rho = \rho_h + (\rho_d-\rho_h)\,\phi, \qquad
  \mu  = \mu_h + (\mu_d-\mu_h)\,\phi. \label{eq:mixing}
\end{equation}

The interface is tracked using the conservative level-set method developed by 
Olsson  et al. ~\cite{olsson2005,olsson2007}. It is represented by the ($\phi=0.5$) contour, with the level-set variable varying smoothly from 0 in one fluid to 1 in the other. The initial profile is defined as 
\begin{equation}
  \phi = \frac{1}{2}\left[1+\tanh\!\left(\frac{d}{2\varepsilon}\right)\right], \label{eq:profile}
\end{equation}
where  $d$ is the signed distance from the interface (positive inside the drop) and $\varepsilon$ controls the interface thickness. 
For the initial circular drop ($t=0$),  $d=a-\sqrt{(x-x_0)^2+(y-y_0)^2}$. 
As the flow evolves, the level-set variable  $\phi$  is advected by the velocity field. A reinitialisation term is included to maintain the interface profile and limit numerical spreading: 
\begin{equation}
  \frac{\partial\phi}{\partial t}+\mathbf{u}\cdot(\nabla\phi)
  = \beta_{ls}\,\nabla\cdot\!\left[\varepsilon\,\nabla\phi
    -\phi(1-\phi)\frac{\nabla\phi}{|\nabla\phi|}\right].
  \label{eq:ls}
\end{equation}
where $\mathbf{u}$ is the velocity field, $\varepsilon$ controls the interface thickness, and $\beta_{ls}$ (m/s) determines the rate at which the interface profile is restored. The right hand side term counteracts numerical spreading and helps maintain a smooth transition between the two fluids as the drop deforms.

The unit normal pointing into the drop, the interface curvature and a smoothed interface delta function are obtained directly from $\phi$:
\begin{equation}
  \mathbf{n}=\frac{\nabla\phi}{|\nabla\phi|}, \qquad
  \kappa=-\nabla\cdot\mathbf{n}, \qquad
  \delta = 6\,\phi(1-\phi)\,|\nabla\phi|. \label{eq:geometry}
\end{equation}
The factor 6 normalises $\delta$ such that $\int\delta\,\mathrm{d}n=\int_0^1 6\phi(1-\phi)\,\mathrm{d}\phi=1$ across the interface.

\subsection{Dimensionless parameters}
\label{sec:nondim}
 
The simulations solve the dimensional governing equations
\eqref{eq:continuity}, \eqref{eq:momentum} and \eqref{eq:ls} using the material
properties listed in Table~\ref{tab:parameters}. The dimensionless parameters used to
organise and report the results are defined below, and their ranges are listed in
Table~\ref{tab:nondim}.

Capillary number Ca (ratio of the viscous stress  to the interfacial stress), Bond number Bo (ratio of gravitational forces arising from the density contrast between the drop and the host magma to interfacial tension), Reynolds number $\mathrm{Re}$ (ratio of inertial and viscous stresses in the host magma) 
are defined as
\begin{equation}
  \mathrm{Ca} = \frac{\mu_h U}{\sigma} ;~~~ \label{eq:Ca}
   \mathrm{Bo} = \frac{\Delta\rho\, g\, a^{2}}{\sigma} ;~~~
  \mathrm{Re} = \frac{\rho_h U H}{\mu_h} .
\end{equation}
The model dimension $L^{*}$ (the aspect ratio of the domain), vertical confinement ratio 
 $C_v$ (ratio of the drop diameter to the height of the chamber), lateral confinement ratio $C_l$ (ratio of the drop diameter to the length of the chamber), Viscosity ratio $\mu_r$ (ratio of the drop to the host viscosity) are expressed as 
 \begin{equation}
  L^{*} = \frac{L}{H}; ~~~   C_v = \frac{2a}{H} ;~~~  C_l = \frac{2a}{L} ;~~~  \mu_r = \frac{\mu_d}{\mu_h}. 
\end{equation}
The dimensionless shear flow timescale  $t^{*}$ is defined as $t^{*} = \dot\gamma\, t$ where shear rate $\dot\gamma = \frac{2U}{H}$. 
The accumulated shear strain ($\gamma$) at time $t$ is $\gamma=\dot\gamma t$ . 
The strain at the instant at which a drop reaches its maximum stretch, $\gamma_{\max}=\dot\gamma\,t_{\max}$.

The drop deformation is characterised by two parameters. The Taylor deformation parameter is  defined as:
\begin{equation}
  D=\frac{l - b}{l + b}, \label{eq:D}
\end{equation}
where $l$ and $b$ are the semi-major and semi-minor axes of the deformed drop, respectively. The parameter $D$ is zero for a circular drop and approaches unity as the drop becomes increasingly elongated. When the shape of the drop is no longer ellipsoidal, the Taylor deformation parameter may not adequately represent the overall elongation. In such cases, the elongation parameter $\delta=l/a$ is used, where $l$ is half the end-to-end length of the drop and $a$ is its initial radius.
The second parameter is the orientation angle, $\theta$, defined as the angle between the drop's major axis and the shear-flow direction.

\begin{table}[htbp]
\centering
\setlength{\tabcolsep}{6pt}
\renewcommand{\arraystretch}{1.2}
\begin{tabular}{|l|p{5cm}|}
\hline
\rowcolor{gray!20}
\textbf{Non-dimensional Parameter} & \textbf{Value range} \\ \hline
Capillary Number ($\text{Ca}$) & $1 \times 10^6$ - $8 \times 10^9$ \\ \hline
Bond Number ($\text{Bo}$) & $3.53 \times 10^6$ and $9.81 \times 10^6$ \\ \hline
Model Dimension ($L^*$) & $1.6$, $2$ and $3$ \\ \hline
Viscosity Ratio ($\mu_r$) & $1 \times 10^{-3}$ - $1 \times 10^{-6}$ \\ \hline
Vertical Confinement Ratio ($2a/H$) & $0.12$ and $0.2$ \\ \hline
Lateral ~Confinement Ratio ($2a/L$) & $4.00 \times 10^{-2}$  -  $1.25 \times 10^{-1}$ \\ \hline
 Reynolds number, $\mathrm{Re}$ & $1.15\times10^{-6}$ - $9.20\times10^{-3}$ \\
\hline
\end{tabular}
\caption{Value range for non-dimensional parameters used to numerically simulate the models.}
\label{tab:nondim}
\end{table}

The following parameters are kept fixed in all simulations: chamber height $H=50$~m,
drop density $\rho_d=2700$~kg\,m$^{-3}$, host density $\rho_h=2300$~kg\,m$^{-3}$
($\Delta\rho=400$~kg\,m$^{-3}$), drop viscosity $\mu_d=10^{3}$~Pa\,s,
interfacial tension $\sigma=10^{-2}$~N\,m$^{-1}$ and gravitational acceleration
$g=9.81$~m\,s$^{-2}$. The varied parameters are listed in Table~\ref{tab:nondim}. 
The host viscosity is varied between $\mu_h=10^{6}$ - $10^{9}$~Pa~s corresponding to $10^{-6} \le \mu_r \le 10^{-3}$. For each $\mu_h$, four wall
speeds $U=0.01$, $0.03$, $0.05$ and $0.08$~m\,s$^{-1}$ are imposed so that
$\mathrm{Ca}=\mu_hU/\sigma$ ranges from $1\times10^{6}$ to $8\times10^{9}$. 
The drop radius ($a=3$ or $5$~m) sets $2a/H$ and $\mathrm{Bo}$, and the chamber length
($L=80$, $100$ or $150$~m) sets $L^{*}$ and $2a/L$ as mentioned in Table~\ref{tab:nondim}.  The 96 simulations cover all combinations of the four $\mu_h$, four $U$, two $a$ and three $L$ values, and are listed in the Appendix~\ref{sec:appendix1}
(Tables~\ref{tab:sim_inputs} and~\ref{tab:sim_outputs}).

\subsection{Numerical implementation and convergence study}
\label{sec:impl}

The governing equations \eqref{eq:continuity}, \eqref{eq:momentum} and \eqref{eq:ls} are solved using the finite element method coupled with the level-set method to
track the interface between the drop and the host magma. The computational domain is discretised using a mesh of triangular and quadrilateral elements, with local refinement around the drop to resolve the interface and its deformation. 
 The level-set variable distinguishes the drop from the surrounding fluid, and the interface is identified by the contour $\phi=0.5$. Its evolution is governed by Eq.~\eqref{eq:ls}, which describes interface advection and maintains a finite interface thickness. The time-dependent equations are integrated using an implicit backward differentiation formula and  an adaptive time step controlled by a relative tolerance of
0.005. 

\begin{table}[htbp]
\centering
\footnotesize
\setlength{\tabcolsep}{5pt}
\renewcommand{\arraystretch}{1.2}
\begin{tabular}{|l|c|c|c|c|}
\hline
\rowcolor{gray!20}
\textbf{Parameter} & \textbf{Normal mesh} & \textbf{Fine mesh} & \textbf{Finer mesh} & \textbf{Extra fine mesh} \\ \hline
Mesh vertices & 2439 & 3583 & 13655 & 49007 \\ \hline
Triangles & 4126 & 6234 & 25308 & 94122 \\ \hline
Quads & 300 & 372 & 800 & 1556 \\ \hline
Edge elements & 174 & 214 & 432 & 834 \\ \hline
Vertex elements & 8 & 8 & 8 & 8 \\ \hline
Number of elements & 4426 & 6606 & 26108 & 95678 \\ \hline
Minimum element quality & 0.3179 & 0.3239 & 0.3333 & 0.3179 \\ \hline
Average element quality & 0.8237 & 0.8452 & 0.9034 & 0.9350 \\ \hline
Element area ratio & 0.05248 & 0.05798 & 0.1528 & 0.1611 \\ \hline
Mesh area ($\text{m}^2$) & 4000 & 4000 & 4000 & 4000 \\ \hline
Fall time to 17.5~m depth (s) & 1965.5 & 1655.2 & 1344.8 & 1344.8 \\ \hline
Total computational time (s) & 57 & 56 & 113 & 412 \\ \hline
\end{tabular}
\caption{Mesh statistics and convergence metrics for the four refinement levels tested.}
\label{tab:mesh}
\end{table}

A grid-convergence test is performed for Sim~2 (see Table~\ref{tab:sim_inputs}) with input parameters  $\mu_r=10^{-3}$, $2a/H=0.12$, $2a/L=0.075$, $\mathrm{Ca}=3\times10^{6}$, $\mathrm{Ca}/\mathrm{Bo}=0.85$ on four progressively refined meshes (Normal, Fine, Finer and Extra fine). The fall time, defined as the time required for the drop to descend a vertical distance of $17.5\,\mathrm{m}$, is used to assess the effect of mesh resolution.  As shown in Table~\ref{tab:mesh}, the fall time decreases from $1965.5\,\mathrm{s}$ for the Normal mesh to $1655.2\,\mathrm{s}$ for the Fine mesh and $1344.8\,\mathrm{s}$ for the Finer mesh. The Extra fine mesh yields identical results to the Finer mesh to the reported precision. Relative to the Extra fine mesh result, the fall time is approximately $46.2\%$ higher on the Normal mesh and $23.1\%$ higher on the Fine mesh. Although the Extra fine mesh contains $95\,678$ elements and requires $412\,\mathrm{s}$ of computational time, the Finer mesh contains $26\,108$ elements and requires only $113\,\mathrm{s}$. 
The Finer mesh is therefore adopted for the subsequent simulations  reported in Section~\ref{sec:results}. 


\section{Results and Discussion}
\label{sec:results}
\textbf{Enclave Deformation Regimes:} The enclaves undergo deformation due to magma mingling inside dynamic magma chamber. Different shapes and sizes found in nature manifest interplay among various forces applicable to magma chamber settings. Our study recognises different deformation regimes depending on dominant force. We broadly categorised deformed enclaves as Sinking and Non-sinking. Due to the initial boundary condition, a global flow is induced in the chamber which has huge effect on the deformed shape of enclaves. In general they get stretched, some get fragmented, some rotate and  {some resist} the deformation. This section will investigate the relative importance of all the dimensionless parameters like Ca, Bo, Re, $\mu_r$, $2a/H$, $2a/L$ and describe how they guide the ultimate deformation regimes (their interpretation is given in Section 4).

Deformation measures. The shape of each droplet is characterised by the deformation parameter $D$, the second shape measure $\delta$, and the orientation angle $\theta$ of its major axis measured from the flow direction. The strain is $\gamma = \dot{\gamma}t$, where $\dot{\gamma}$ is the imposed shear rate, and the normalised depth ($H^{*}= 2d/H$) is the depth ($d$) of the deformed droplet's centre relative to original centre depth ($H/2$). Break-up is registered when particular areas on droplet body gets thinned and ultimately pinched off.

\subsection{Sinking}

\begin{figure}[!htb]
\begin{center}
\includegraphics[scale=0.5]{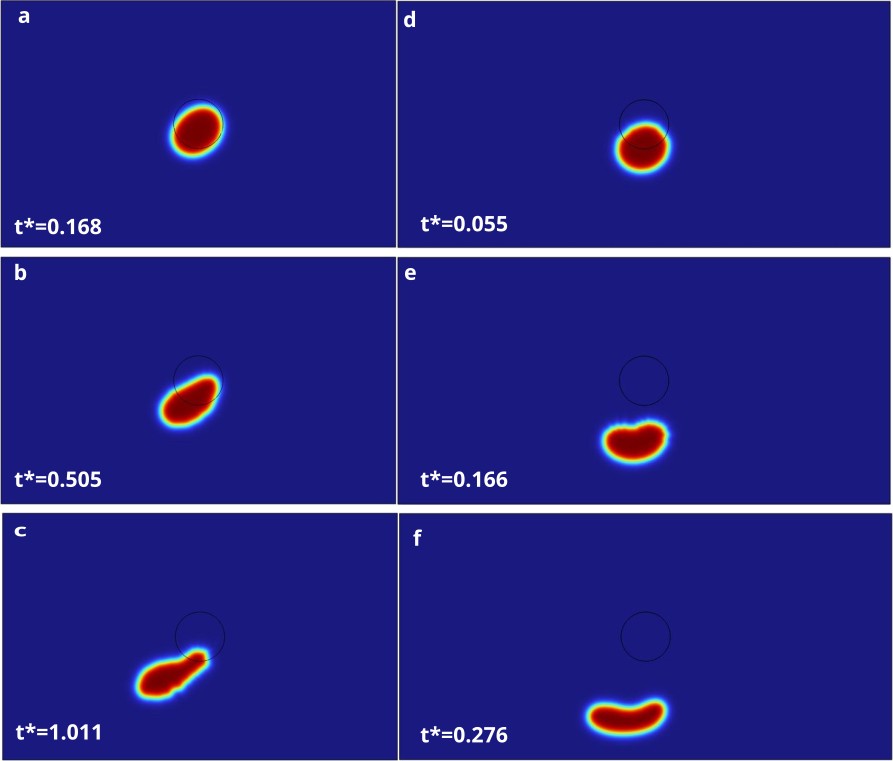}
\caption{Descent and progressive deformation of the magma drop through time for lateral confinement ratios 0.125 and 0.1 (left and right panels respectively).}
\label{fig:fig3}
\end{center}
\end{figure}

We investigate deformation of the dense magma drop in a viscous host with the fixed density contrast ($\Delta\rho = 400$ kg/m$^3$) and higher viscosity ratio ($\mu_r = 10^{-3}$). Fig.~\ref{fig:fig3} presents the progressive change of the drop shape under two different lateral confinement ratios. Left column corresponds to $2a/L = 0.125$ (Sim 5), where Fig.~\ref{fig:fig3}a starts with deformed droplet at $t^* = 0.168$ where it sinks as well as stretches with respect to original undeformed position and shape. The deformation can be measured by the values of $D$, $\delta$ and $\theta$ which are 0.12, 1.11 and 40.90 respectively. Fig.~\ref{fig:fig3}b and c show that the drop continue to sink while getting more deformed with narrower upper right edge when time progressed from $t^* = 0.505$ to $t^* = 1.011$. Corresponding $D$, $\delta$ and $\theta$ values are 0.31, 1.32 and 38.00 for $t^* = 0.505$ and 0.48, 1.57 and 31.18 for $t^* = 1.011$. Ca/Bo for this simulation has been recorded as 0.1.  Fig.~\ref{fig:fig3}d-f shows the pattern of deformation at lower confinement ($2a/L = 0.1$, Sim 13). At $t^* = 0.055$ the droplet starts to sink and deform (Fig.~\ref{fig:fig3}d), with $D = 0.05$ and $\delta = 1.04$. At $t^* = 0.166$, the droplet deforms further to form an kidney bean shape (Fig.~\ref{fig:fig3}e) as $D$ increased to 0.31 and $\delta = 1.28$. With increasing time ($t^* = 0.276$) the drop keeps descending towards the base and its shape get laterally flattened with $D$ valued at 0.50 and $\delta = 1.57$. However, unlike Sim 5, Sim 13 doesn't get inclined and $\theta$ is maintained at $\sim 0^\circ$. Deformed drop in both cases doesn't sink vertically, rather they shift leftward as they go down. It can also be noticed that the drop reaches the base much faster in Sim 13 than Sim 5. We assess the relative effect of Capillary number (Ca) and Bond number (Bo) and it is found that both have $\text{Ca/Bo} = 0.1$. This low ratios indicate gravitational forces dominate over viscous forces causing heavy magma drop to sink. Further simulations with $\mu_r = 10^{-3}$ yield that their range of Ca/Bo varies from 0.1 to 2.27. The simulations with $\text{Ca/Bo} < 1$ can be grouped as purely \texttt{Sinking enclaves}. The normalized depths ($H^*$) for these sinking enclave drops are 0.7 and 0.23 respectively.

\subsection{Non-sinking}
We differentiate Non-sinking enclaves from Sinking based on the values of $Ca/Bo$. The cases where $\text{Ca/Bo} > 1$, can be categorised as Non-sinking as gravitational forces are relatively weaker for those simulations. The Non-sinking deformation can again be broadly classified among three deformation processes - a) stretching-and-breakup, b) stretching-and-rotation and, c) stretching-and-retraction. All these enclaves essentially undergo initial tilt and elongation. However eventually the deformation pattern diversify depending on selected non-dimensional parameter settings and their combinations.

\subsubsection{Stretching with Breakup}

We simulated models using large range of viscosity ratio ($\mu_r = 10^{-6}-10^{-3}$) where in all cases host is more viscous than the droplet. It is found that decreasing viscosity ratio has a positive impact on droplet stretching. Intense stretching causes steady increase in long axis and decrease in short axis dimension, ultimately causing formation of neck along which the drop get fragmented. We also use the Capillary number ($\text{Ca}$), lateral confinement ratio ($2a/L$) and vertical confinement ratio ($2a/H$) to understand their effect on droplet breakup. Among the 96 simulations we performed for this study, this mode of deformation is the most widely observed outcome.

Breakup occurs with the stretching of the magma droplets to their maximum limit, and a pinch-off point is formed at which the magmatic droplet breaks apart. Depending on the preferred sites for pinch-off, we can identify three kinds of droplet break-up which finally controls the daughter droplet numbers and sizes - i) Edge break-up, ii) Mid break-up, iii) Multi break-up. The types are identified by the first break-up of a particular droplet. If there is only one preferential site for pinch off, there will be a single break-up at first. We find single pinch off locations generally forms either near top-right edge or near the middle of stretched droplet. Whereas, multiple pinch off locations are caused by thinning at more than one location along the long axis of the elongated and deformed droplet. Eventually these secondary droplets may undergo further break-ups, which are not considered during break-up nomenclature. We will discuss the individual deformation patterns of these three as well as their transitions driven by change in parameter settings. 

At viscosity ratio between drop and host decreases further ($\mu_r = 10^{-4}$) from Sinking regime ($\mu_r = 10^{-3}$), we encounter Stretching with Break-up. Figure 4 shows the transient deformation history for Sims 34 (left panel) and 38 (right panel) where first can be identified as Mid break-up and second as Edge break-up. We present the results upto the time of pinch-off locations are fixed.

\begin{figure}[!htb]
\begin{center}
\includegraphics[scale=0.5]{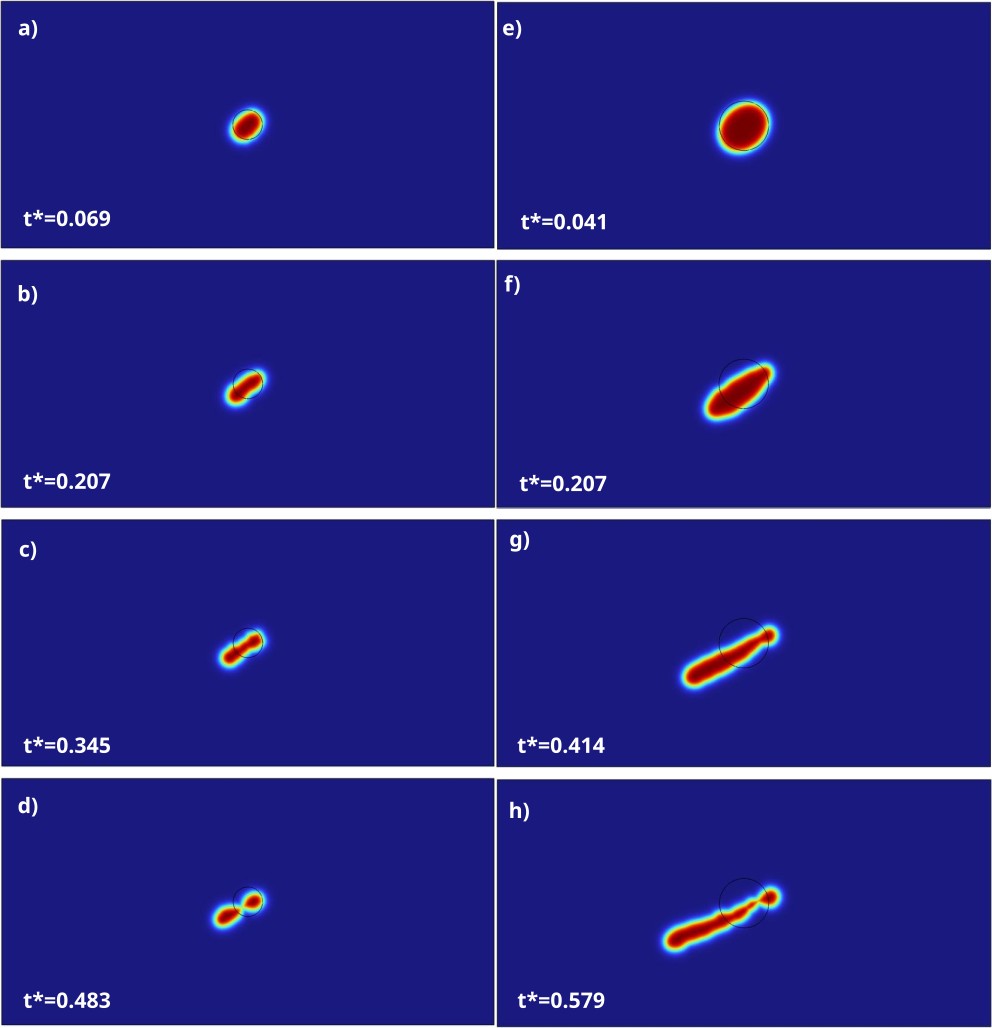}
\caption{Pinching and breakup mechanism at mid-point (left panel) and at edge (right panel) for lateral dimension ratio $0.1$, vertical confinement ratios of $0.12$ and $0.2$, viscosity ratio $1.00 \times 10^{-4}$, $\text{Ca} = 3.00 \times 10^7$.}
\label{fig:fig4}
\end{center}
\end{figure}

Sim 34 has been performed at $\text{Ca} = 3.00 \times 10^7$ and $\text{Re} = 3.45 \times 10^{-4}$. In Figure~\ref{fig:fig4}a, we see the initial deformation of the magma droplet at $t^* = 0.069$. The droplet deforms into an ellipse having $\delta = 1.16$ and $D = 0.20$. $\theta$ value at this stage is 43.52 which is changed to 38.72 as time progresses to $t^* = 0.207$ (Figure~\ref{fig:fig4}b). Corresponding $\delta$ and $D$ are increased to 1.45 and 0.43 respectively which indicates higher deformation due to stretching. At $t^* = 0.345$, the magma droplet stretches further (Figure~\ref{fig:fig4}c) causing higher $D$ (0.57) and $\delta$ (1.59), whereas $\theta$ shows 33.94. Finally pinch off is observed at $t^* = 0.483$ (Figure~\ref{fig:fig4}d) when the droplet attains its maximum stretch as a continuous body ($\delta = 1.78$). Thickness is minimum at this stage with $D = 0.96$. The site of the pinch off is located near the middle of the long axis, making this case a representative of mid break-up. The $\text{Ca/Bo}$ in this case is 8.49, vertical and lateral confinement ratios are 0.12 and 0.06 respectively.

A transition occurs when $\text{Ca/Bo}$ reduces to 3.06 and vertical \& lateral confinement ratios are changed to 0.2 and 0.1 respectively. This corresponds new trend of deformation (Figures~\ref{fig:fig4}e-h) in Sim 38 even though the $\text{Ca}$ and the $\text{Re}$ values are the same as Sim 34. Figure~\ref{fig:fig4}e displays early deformation at $t^* = 0.041$ where an ellipse is formed. The value of the deformation parameter is relatively low ($D = 0.12$) and corresponding $\delta$ is calculated to be 1.11. At $t^* = 0.207$, the droplet stretches into a deformed ellipsoid (Figure~\ref{fig:fig4}f) of dimensions with increased values $D$ and $\delta$ of 0.49 and 1.60. At the same time the deformed magma drop has a tilt ($\theta$) of 36.68. At $t^* = 0.414$, the deformation looks asymmetric where top-right corner is narrower than the rest of the body (Figure~\ref{fig:fig4}g). Corresponding  $D$, $\delta$ and $\theta$  values are 0.73, 2.11 and 29.95 respectively. This asymmetry increases with time ($t^* = 0.579$) and prominent pinch off is formed near right edge (Figure~\ref{fig:fig4}h). At this stage $D$ increases further to $D = 0.92$ and $\delta$ to 2.51, whereas $\theta$ reduces to 24.66. Though the locations of break-up are different in these two cases, both of them belong to Non-sinking regime as $\text{Ca/Bo} > 1$. Time taken to get stretched and pinched is relatively more in Sim 38 than Sim 34. Interestingly, we still can see effect of body force in Sim 38 as the centre of the deformed drop lies beneath the centreline of the chamber ($H^* = 0.88$).

\begin{figure}[!htb]
\begin{center}
\includegraphics[scale=0.5]{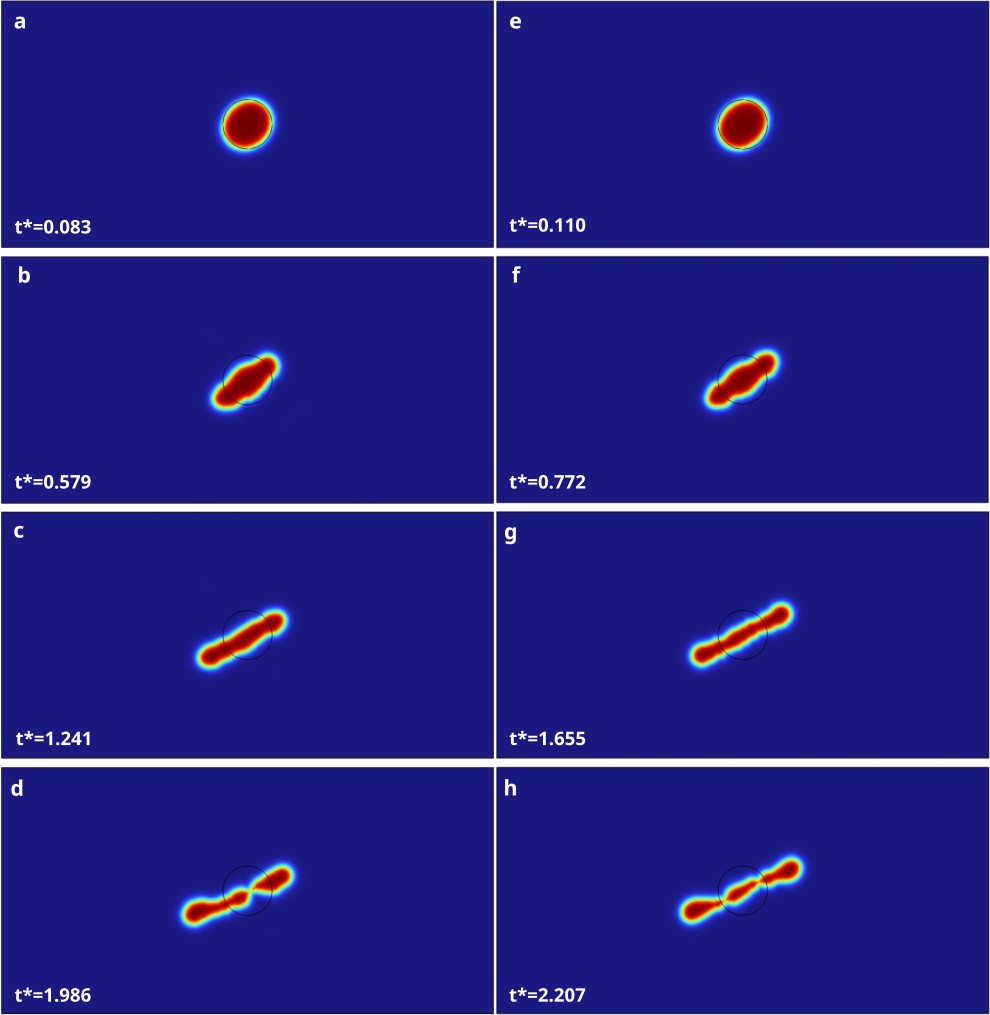}
\caption{Mid-point breakup (left panel) and multiple breakup (right panel) mechanism of the magma droplet under shear for lateral confinement ratio $0.1$; vertical confinement ratio $0.2$, viscosity ratio $1.0 \times 10^{-5}$ and $1.0 \times 10^{-6}$, shear rate $4.00 \times 10^{-4}$, $\text{Ca} = 1.0 \times 10^8$ and $1.0 \times 10^9$.}
\label{fig:fig5}
\end{center}
\end{figure}

Within Non-sinking regime, single pinch off leading to either edge or mid break-up spans almost all over the total viscosity range of current study i.e. $\mu_r = 10^{-6}-10^{-3}$. However, occurrence of multiple pinch off causing multi break-up is generally found at lower viscosity range ($\mu_r = 10^{-6}-10^{-5}$). In {Fig.~\ref{fig:fig5}} we will present a mid-break-up case (Sim 61) at $\mu_r = 10^{-5}$ and its transition to multi break-up (Sim 85) at $\mu_r = 10^{-6}$. Corresponding $\text{Ca/Bo}$ increase from 10.2 to 102 from Sim 61 to Sim 85. In both simulations, the vertical confinement ratio and the lateral confinement ratio were maintained at 0.2 and 0.1.

In Fig.~\ref{fig:fig5}a, we see that the magma droplet undergoes mild deformation at $t^* = 0.083$ for Sim 61. Deformation related parameters are measured to be $D = 0.11$, $\delta = 1.07$. Fig.~\ref{fig:fig5}e represents equivalent initial deformation for Sim 85 at $t^* = 0.110$. The deformed shapes at this preliminary stage are almost similar for both cases with later having slightly higher $D$ ($= 0.13$) and $\delta$ ($1.10$). The droplet continues to deform with an increase in time, and at the timestep $t^* = 0.579$, it deforms to take the shape of a distorted ellipse (Fig.~\ref{fig:fig5}b) with $D$ ($= 0.52$) and $\delta$ ($=1.55$). Though the equivalent stage for Sim 85 at $t^* = 0.772$ shows same pattern of deformation (Fig.~\ref{fig:fig5}f), it shows higher stretching. The corresponding values of $D$ and $\delta$ are 0.54 and 1.67. If we compare the $\delta$ and $\theta$ values, we find that $\delta$ increases from 1.96 in Sim 61 to 2.23 in Sim 85, whereas $\theta$ is 28.91 in Sim 61 and 27.71 in Sim 85. Drop in Sim 85 continue to stretch more than that in Sim 61 as time progresses (Fig.~\ref{fig:fig5}c \& g). Finally, single pinch off occurs in Sim 61 at $t^* = 1.986$ with $D = 0.95$, $\delta = 2.37$ and $\theta = 24.55$. Similarly, in Sim 85, at $t^* = 2.207$, with further stretching and deformation ($D = 0.93$), necking of droplet occurs at two different positions (Fig.~\ref{fig:fig5}h). This resultant multi break-up enclave shape looks like a tilted ``candy'' having $\theta = 23.72$.

\subsubsection{Stretching with Rotation}

Another deformation pattern of magma droplet is observed in our simulation where the droplet undergoes high strain and rotates toward the applied shear direction. In some cases, the rotation continues beyond this alignment and the stretched droplet enters the zone of compression crossing the zone of extension. The resultant deformed droplet resists further stretching and break-up and is folded instead. We notice that this type of deformation occurs when the confinement ratios are relatively higher ($2a/H = 0.2$ and $2a/L = 0.125$) among the simulations. However, this parameter alone can't guide the deformation as there are many break-up cases in current study also have same parameter ratios. We choose Sim 77 and Sim 80 and observe what change can cause the transition from stretching with break-up to stretching with rotation.

\begin{figure}[!htb]
\begin{center}
\includegraphics[scale=0.5]{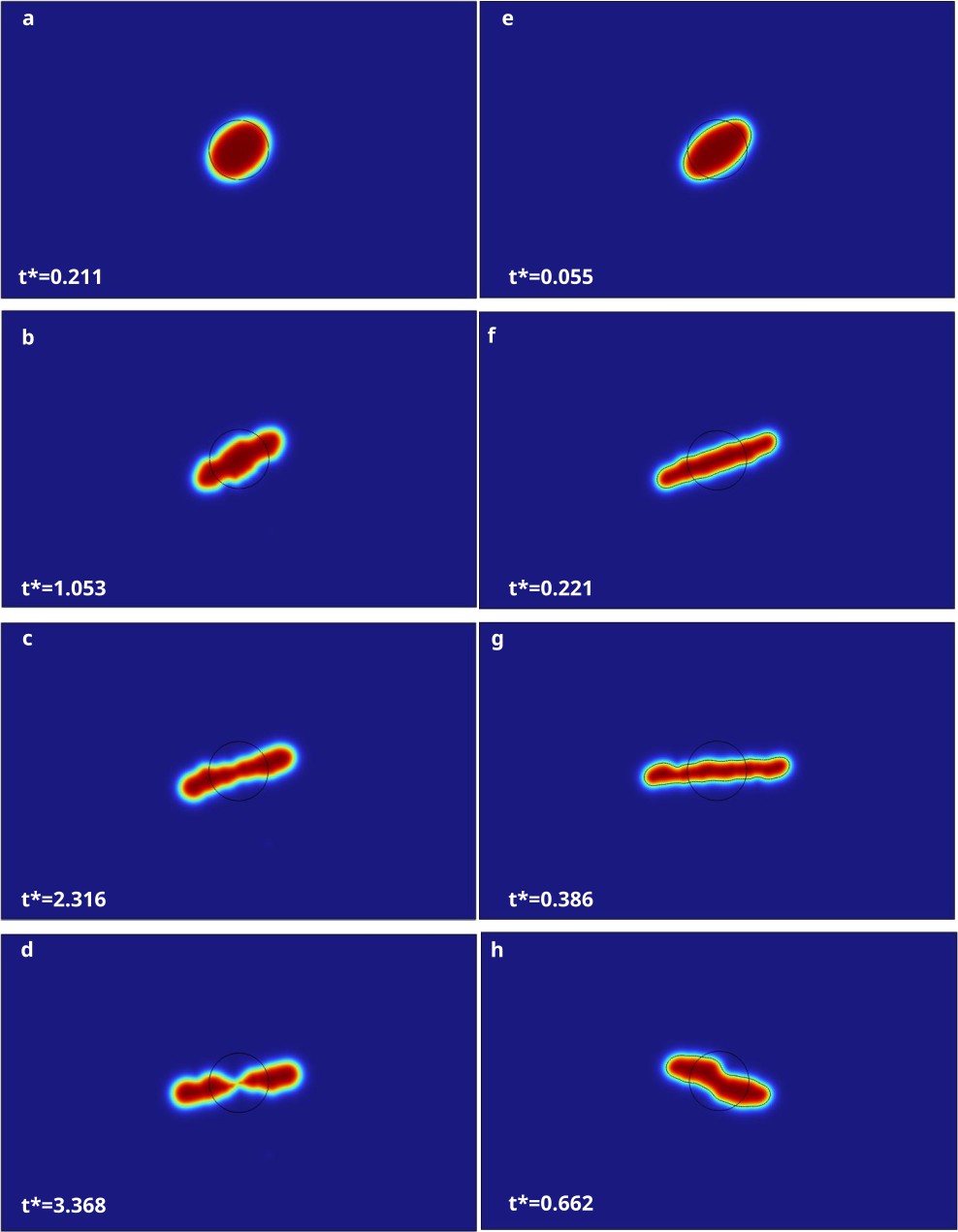}
\caption{Mid-point breakup (left panel) and Stretching with Rotation (right panel) mechanism at lateral confinement ratio $0.125$; vertical confinement ratio $0.2$, viscosity ratio  $1.0 \times 10^{-6}$, shear rates $4.00 \times 10^{-4}$ and $3.20 \times 10^{-3}$ , $\text{Ca} = 1.0 \times 10^9$ and $8.0 \times 10^9$.}
\label{fig:fig6}
\end{center}
\end{figure}

In Fig.~\ref{fig:fig6}a, we see the initial deformation of the magma droplet at $t^* = 0.211$ for Sim 77. The droplet deformation was measured and the corresponding values of $D$, $\delta$ and $\theta$ were $0.17$, $1.14$ and $43.40$ respectively. Fig.~\ref{fig:fig6}e depicts initial deformation for Sim 80 at $t^* = 0.055$ with $D$ value of $0.31$ which is higher than the former simulation. The $\delta$ and $\theta$ values are $1.31$ and $37.46$. At $t^* = 1.053$, the droplet deforms to take the shape of an asymmetric ellipse with $D = 0.53$, $\delta = 1.66$ and $\theta$ valued at $30.03$ as shown in Fig.~\ref{fig:fig6}b. Fig.~\ref{fig:fig6}f depicts the next stage deformation for sim 80, wherein the ellipse is stretched to form an oblate disc at $t^* = 0.221$. The deformation parameters $D$, $\delta$ and $\theta$ for this stage are valued at $0.68$, $2.13$ and $20.78$ respectively. Fig.~\ref{fig:fig6}c shows the deformation for sim 77 where the ellipse is stretched and the corresponding values of $D$, $\delta$ and $\theta$ are $0.65$, $2.01$ and $18.90$ at $t^* = 2.316$. Fig.~\ref{fig:fig6}g shows the equivalent stage of deformation for sim 80 at $t^* = 0.386$. The oblate droplet is further stretched and the $D$ and $\delta$ values measure $0.78$ and $2.45$. The inclination of the droplet reduces to $5.01$ and appears horizontal. Fig.~\ref{fig:fig6}d shows the final pinch off stage for Sim 77 at $t^* = 3.368$. The stretched magma droplet further stretches to form a narrow pinch off point from which it breaks off with $D$, $\delta$ and $\theta$ measuring $0.93$, $2.17$ and $11.39$. Fig.~\ref{fig:fig6}h shows that the magma droplet resists further deformation ($D = 0.64$), and folds to form a faint ``z'' shape at $t^* = 0.662$.

\subsubsection{Stretching with Retraction}

\begin{figure}[!htb]
\begin{center}
\includegraphics[scale=0.54]{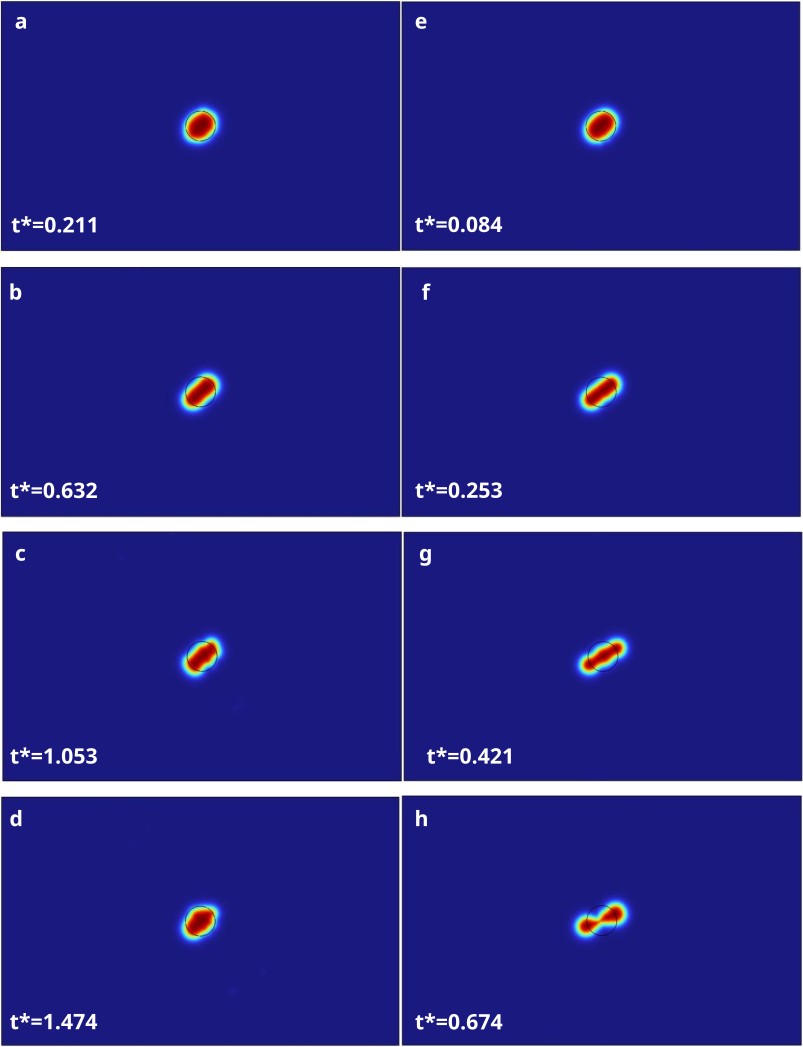}
\caption{Stretching with retraction (left panel) and Mid-point breakup (right panel) mechanism at lateral confinement ratio $0.075$; vertical confinement ratio $0.12$, viscosity ratio $1.0 \times 10^{-6}$, shear rates $4.00 \times 10^{-4}$ and $1.20 \times 10^{-3}$, $\text{Ca} = 1.0 \times 10^9$ and $3.0 \times 10^9$.}
\label{fig:fig7}
\end{center}
\end{figure}

We observe another deformation pattern of the magma droplet in our simulations, where the droplet under the effect of low strain undergoes little stretching as compared to other cases and then compress or retracts. We notice this type of deformation at lower vertical confinement ratio ($2a/H = 0.12$) and over a range of vertical confinement ratios ($2a/$L = 0.075 - 0.04). We choose Sim 73 and 74 for the study of the deformational pattern for the transition of stretching with breakup and stretching with retraction and what changes in parameters act as the driving force behind this transition.

The transition from stretching with retraction to stretching with breakup is observed at lower viscosity ratio ($10^{-6}$). Increase in velocity contributes to the increase in the capillary number ($1.0 \times 10^{9}$ - 3.0$ \times 10^{9}$) which in turn promotes the transition. In {Fig.~\ref{fig:fig7}} we present a stretching and retraction case (Sim 73) having $\text{Ca} = 1.00 \times 10^{9}$ and a mid breakup case (Sim 74) with $\text{Ca} = 3.00 \times 10^{9}$. The corresponding $\text{Ca/Bo}$ value increases from 283 to 849.

In Fig.~\ref{fig:fig7}a, the magma droplet initially deforms at $t^* = 0.211$ for Sim 73. The deformation parameters $D$, $\delta$ and $\theta$ were measured to be $0.16$, $1.12$ and $42.94$ respectively. Fig.~\ref{fig:fig7}e represents the initial deformation of magma droplet for Sim 74 at $t^* = 0.084$ where the $D$, $\delta$ and $\theta$ have the corresponding value of $0.19$, $1.16$ and $42.81$. Fig.~\ref{fig:fig7}b shows the deformation at $t^* = 0.632$ for Sim 73, where the magma droplet is stretched with $D$ valued at $0.35$, which is higher than the earlier timestep. The corresponding values of $\delta$ and $\theta$ are $1.31$ and $41.03$. The equivalent step for Sim 74 is shown by Fig.~\ref{fig:fig7}f, in which the magma droplet undergoes more stretching, depicted by $D = 0.41$ at $t^* = 0.253$. The corresponding values of $\delta$ and $\theta$ are $1.42$ and $36.44$, respectively. Fig.~\ref{fig:fig7}c shows the deformation at $t^* = 1.053$ where the magma droplet, instead of experiencing more deformation, resists it and starts to retract. The value of $D$ and $\delta$ does not change for this step while the value of $\theta$ is decreased to $40.74$. Fig.~\ref{fig:fig7}g shows the further deformation of the magma droplet at $t^* = 0.421$. The droplet is stretched and forms an ellipse with uneven boundary and the value of $D$, $\delta$ and $\theta$ are $0.53$, $1.60$ and $31.07$, respectively. Fig.~\ref{fig:fig7}d shows the retracted magma droplet at $t^* = 1.474$ with a reduction in the values of $D$, $\delta$ and $\theta$ to $0.28$, $1.25$ and $40.06$. Fig.~\ref{fig:fig7}h shows the mid-point pinching of the magma droplet due to further deformation for Sim 74. The corresponding values of $D$, $\delta$ and $\theta$ are $0.77$, $1.67$ and $24.16$.

\subsection{Quantitative synthesis of deformation and regime selection}

\begin{figure}[!htb]
\begin{center}
\includegraphics[scale=0.5]{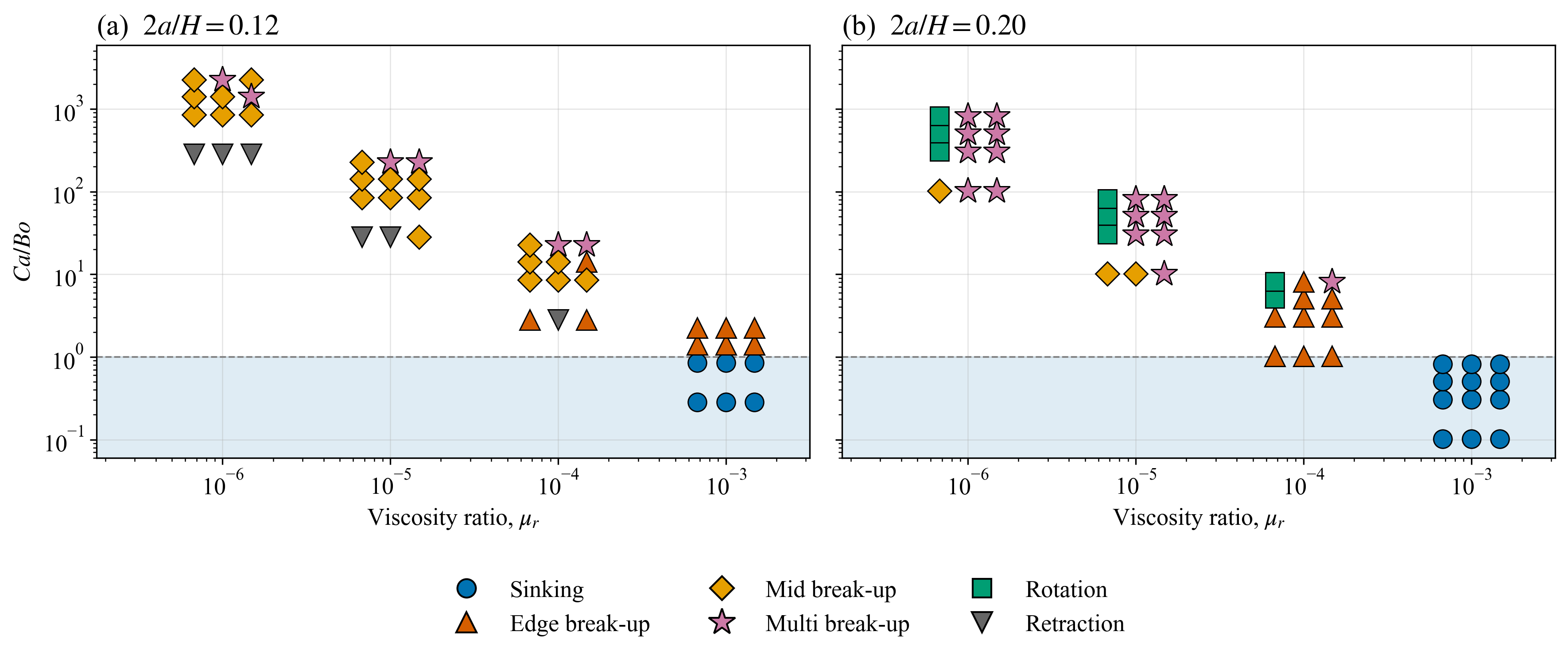}
\caption{Deformation-regime map in the $\text{Ca/Bo} $-$\mu_r$ parameter space for $2a/H = 0.12$ and $0.20$. Symbols identify sinking, edge break-up, mid break-up, multi break-up, rotation, and retraction.}
\label{fig:fig8}
\end{center}
\end{figure}

\begin{figure}[!htb]
\begin{center}
\includegraphics[scale=0.69]{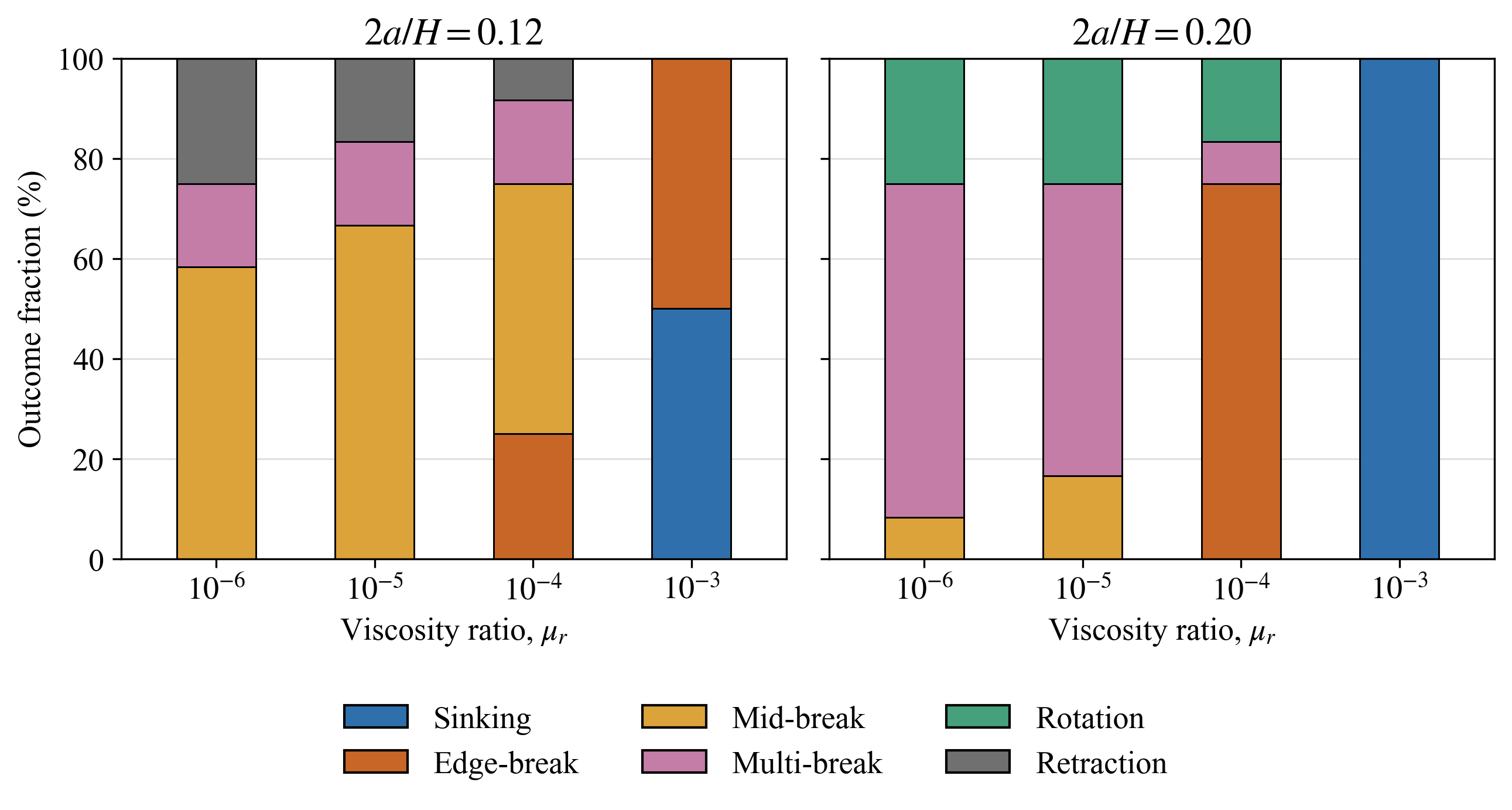}
\caption{Approximate outcome fractions as a function of viscosity ratio, $\mu_r$, for the two vertical confinement ratios. Increasing $\mu_r$ shifts the outcome distribution toward sinking, while lower $\mu_r$ favours break-up and, at higher confinement, rotation.}
\label{fig:fig9}
\end{center}
\end{figure}

\begin{figure}[!htb]
\begin{center}
\includegraphics[scale=0.59]{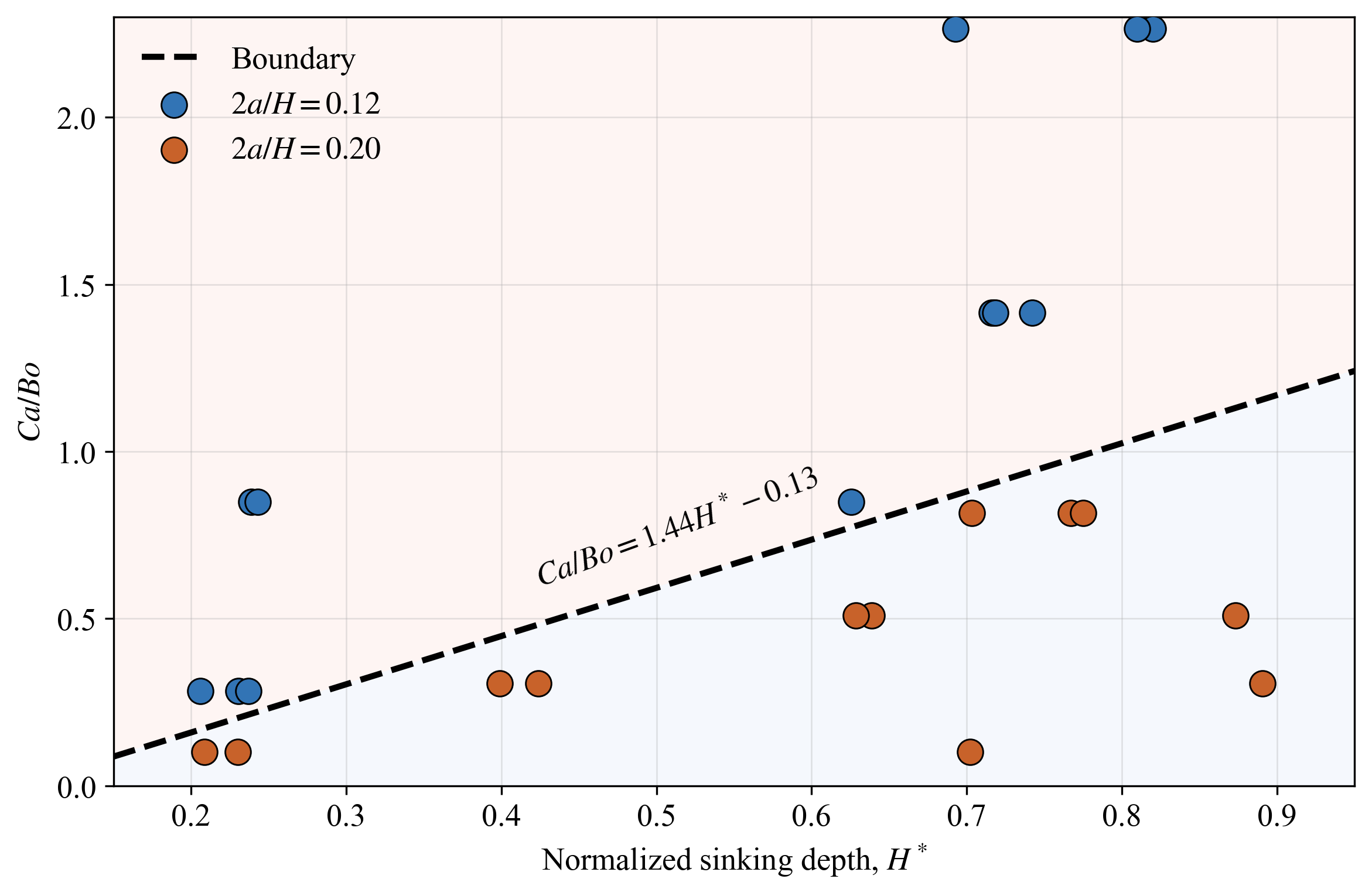}
\caption{$\text{Ca/Bo}$ as a function of normalized sinking depth, $H^*$, for the two vertical confinement ratios. The dashed curve marks the empirical transition between the gravity-dominated and shear-dominated regions represented by the simulations.}
\label{fig:fig10}
\end{center}
\end{figure}

This section quantifies the deformation patterns described in Section 3.1 across the full set of 96 simulations. Section 3.2 presents the regime map, the outcome statistics and the sinking transition (Figs.~\ref{fig:fig8}-\ref{fig:fig10}), and Section 3.3 presents the orientation and strain at maximum stretch of the non-sinking enclaves (Figs.~\ref{fig:fig11} and \ref{fig:fig12}).

Quantitative analysis is very necessary to understand and relate the deformation patterns and the force balances behind them. Here we will try to synthesize the transient history of deformation discussed in section 3.2.

Fig.~\ref{fig:fig8} maps the $\text{Ca/Bo}$ values of all simulations against the viscosity ratio $\mu_r$ and summarises the deformation regimes in the $\text{Ca/Bo}-\mu_r$ parameter space. The sinking and non-sinking regimes separate near $\text{Ca/Bo} = 1$: sinking occurs predominantly below this value and non-sinking above it. Within the non-sinking field the simulations span a broad range of $\text{Ca/Bo}$ and $\mu_r$, and three sub-regimes are identified: stretching with break-up, stretching with rotation and stretching with retraction. All of them are stretched initially, as they lie at high $\text{Ca/Bo}$. Fig.~\ref{fig:fig8}a shows the results for $2a/H = 0.12$ and Fig.~\ref{fig:fig8}b those for $2a/H = 0.2$. Lower $\mu_r$ is associated with more intense deformation. Stretching with rotation and stretching with retraction are mutually exclusive: retraction is found at low confinement and rotation at high confinement.

Fig.~\ref{fig:fig9} complements Fig.~\ref{fig:fig8} by quantifying the regime distribution as fractions of outcomes. At $2a/H = 0.2$, the high-$\mu_r$ range is 100\% sinking, whereas at $2a/H = 0.12$ the outcomes are shared equally between sinking and edge break-up. Mid break-up is the main outcome over the mid-to-low $\mu_r$ range at $2a/H = 0.2$, with retraction and multi break-up next in importance. At $2a/H = 0.2$, multi break-up is prominent at $\mu_r = 10^{-6}$-$10^{-5}$, where rotation is also observed, and edge break-up becomes significant around $\mu_r = 10^{-4}$. These fractions reflect how the 96 simulations were distributed across the parameter space and should not be read as probabilities of occurrence.

Because $\text{Ca/Bo}$ represents the relative forcing of shear and gravitational buoyancy, it provides a measure of the sinking transition that does not rely on the regime classification. Fig.~\ref{fig:fig10} shows the normalized depth $H^*$ as a function of $\text{Ca/Bo}$ at high viscosity ratio ($\mu_r = 10^{-3}$). Simulations with low $\text{Ca/Bo}$ sink as deep as $H^* = 0.2$, whereas relatively large $\text{Ca/Bo}$ values are associated with shallower displacement. The lower-confinement simulations ($2a/H = 0.12$) cluster at low $\text{Ca/Bo}$ ($< 1$), in contrast to those at higher confinement ($2a/H = 0.2$). The transition therefore occurs over a range of $\text{Ca/Bo}$ that depends on confinement, and not at a single threshold.

\subsection{Orientation and strain at maximum stretch}

\begin{figure}[!htb]
\begin{center}
\includegraphics[scale=0.52]{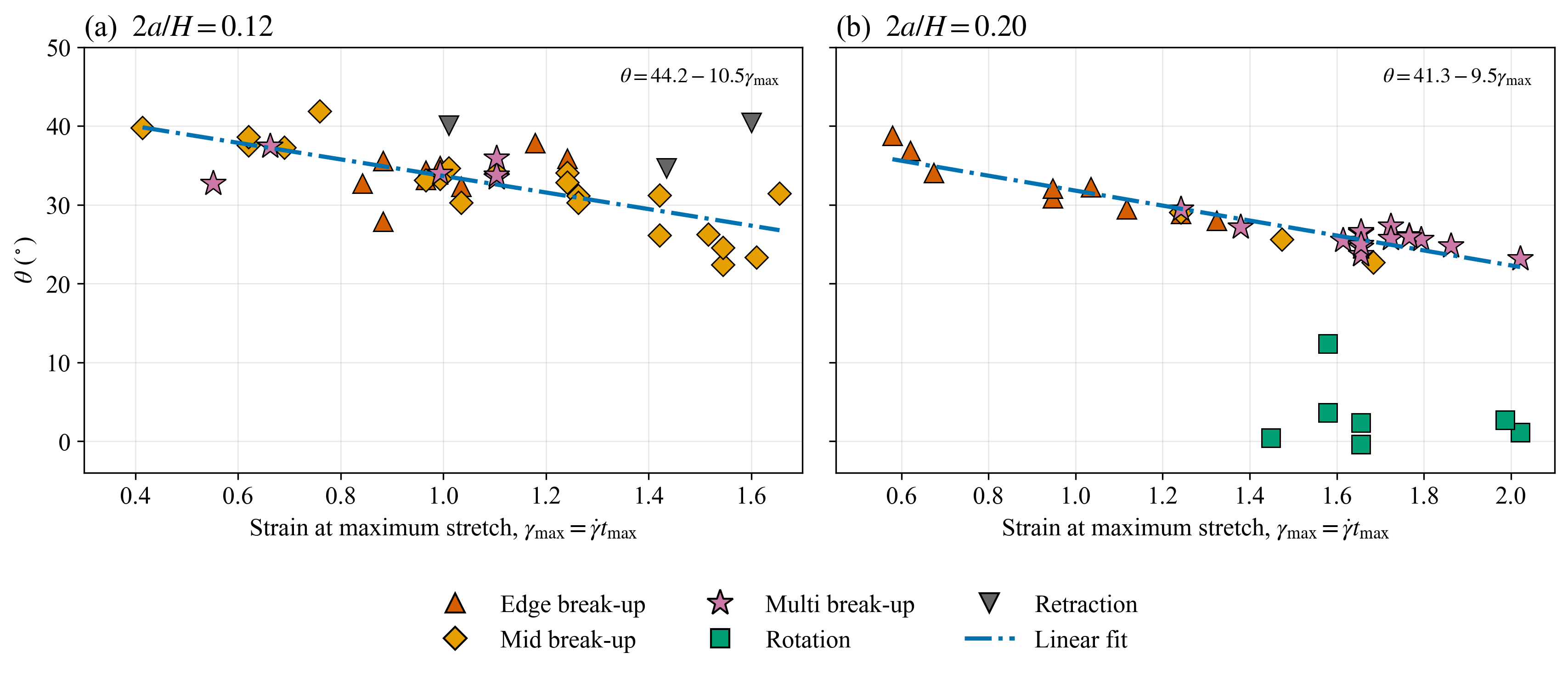}
\caption{Orientation angle, $\theta$, versus strain at maximum stretch, $\gamma$, for $2a/H = 0.12$ and $0.20$. The dash-dotted lines are major-axis linear fits.}
\label{fig:fig11}
\end{center}
\end{figure}

For the non-sinking enclaves we examine the relation between the orientation angle $\theta$ and the strain at the stage of maximum stretching ($\gamma_{\max}$). In our simulations, just after this stage the droplet pinches off and breaks up. $\theta$ decreases linearly with $\gamma$ for both confinements (Figs.~\ref{fig:fig11}a and b), indicating progressive reorientation of the major axis as deformation accumulates. The fitted relations are $\theta = 44.2 - 10.5\gamma_{\max}$ for $2a/H = 0.12$ ($R^2 = 0.54$) and $\theta = 41.3 - 9.5\gamma_{\max}$ for $2a/H = 0.20$ ($R^2 = 0.88$). The slopes are comparable, suggesting that the rate of reorientation is broadly similar for the two confinement ratios, although the difference in $R^2$ indicates that the linear relationship is more consistent for $2a/H = 0.20$.  For the lower confinement ratio, the data show greater scatter, particularly at intermediate and higher strains. Overall, high confinement simulations accumulate high strain. Rotation regime $\gamma_{\max}$ lies at higher end but orientation angle is characterically low  Fig.~\ref{fig:fig11}b which distinguish it from the other deformation modes. 


\begin{figure}[!htb]
\begin{center}
\includegraphics[scale=0.52]{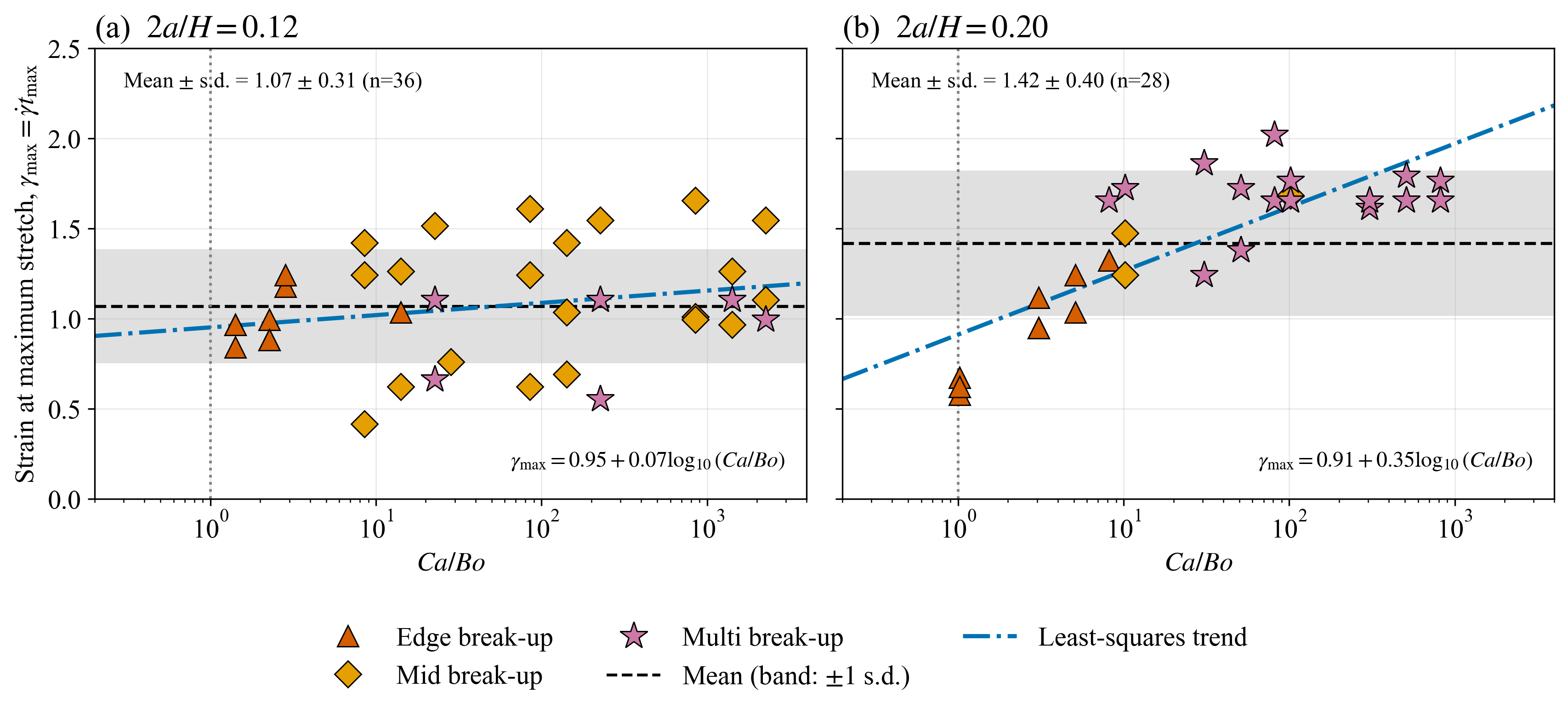}
\caption{Strain at maximum stretch, $\gamma = \dot{\gamma}t_{\max}$, as a function of $\text{Ca/Bo}$ for vertical confinement ratios $2a/H = 0.12$ and $0.20$. The dashed line and shaded band denote the panel mean and $\pm 1$ s.d.; the dash-dotted line is the least-squares trend.}
\label{fig:fig12}
\end{center}
\end{figure}

Among the non-sinking simulations, break-up cases dominate, ranging from edge to mid to multi break-up. Fig.~\ref{fig:fig12} shows the strain accumulated before maximum stretch, $\gamma_{\max} = \dot{\gamma}t_{\max}$, as a function of $\text{Ca/Bo}$. $\gamma_{\max}$ increases with $\text{Ca/Bo}$ for both confinements. For $2a/H = 0.12$ the fitted trend is $\gamma_{\max} = 0.95 + 0.07 \log_{10}(\text{Ca/Bo})$, and for $2a/H = 0.20$ it is $\gamma_{\max} = 0.91 + 0.35 \log_{10}(\text{Ca/Bo})$. The panel means are $1.07 \pm 0.31$ and $1.42 \pm 0.40$, respectively, so the dependence of $\gamma_{\max}$ on $\text{Ca/Bo}$ is stronger at the higher vertical confinement. The slope for $2a/H = 0.12$ is small compared with the scatter about the mean, so the contrast between the two confinements should be treated as indicative.


\section{Discussion}
\label{sec:discussions}

The simulations show how a dense, low-viscosity droplet in a viscous host responds to the competition between shear and buoyancy. Figs.~\ref{fig:fig3}-\ref{fig:fig12} together form an ensemble which consistently provides a fluid-dynamic basis for interpreting the diversity of natural enclave shapes. We discuss the controls on regime selection, the regime-forming mechanisms, the implementation of droplet dynamics for natural enclaves as the novelty of the study and the limitations of our model.

\subsection{Controls and Mechanisms behind the regimes}

$\text{Ca/Bo}$ is the first-order control on whether an enclave sinks or is deformed by the host flow. At higher $\text{Ca/Bo}$ the shear-induced flow overpowers gravity-driven motion and suppresses sinking (Figs.~\ref{fig:fig8} and \ref{fig:fig10}). Because the boundary occurs over a range of $\text{Ca/Bo}$ that depends on confinement (Fig.~\ref{fig:fig10}), it is better viewed as a transition zone than as a sharp boundary at $\text{Ca/Bo} = 1$. Within the non-sinking field, lower viscosity ratio promotes stretching, because a highly viscous host stretches and aligns the low-viscosity enclave more readily, and multiple pinch-off appears only at the lowest ratios ($\mu_r = 10^{-6}-10^{-5}$). However, the viscosity ratio alone does not fix the final morphology within this field. Confinement ($2a/H$) and shear intensity ($\text{Ca/Bo}$) determine how the available strain is accommodated, whether by pinch-off, rotation or retraction (Fig.~\ref{fig:fig9}), and together they distribute the accumulated strain to reshape and reorient the droplet after stretching instead of causing immediate pinch-off.
At the Capillary numbers explored here ($\text{Ca} \gg 1$), interfacial tension has negligible contribution, so the deformation histories described in Section 3 are set by the competition between the host flow, which stretches and rotates the droplet, and buoyancy, which displaces it. In the sinking regime ($\text{Ca/Bo} < 1$) the droplet descends through the host while the imposed global flow advects it sideways (Fig.~\ref{fig:fig3}). For $\text{Ca/Bo} > 1$ the droplet is rotated by the host flow almost similarly in both confiments, while the viscosity ratio determines how freely it stretches.The location of the first pinch-off appears to reflect how strongly buoyancy perturbs the otherwise symmetric stretching. In Sim 34 ($\text{Ca/Bo} = 8.49$) the droplet thins near the middle of its long axis, whereas in Sim 38 ($\text{Ca/Bo} = 3.06$), at the same $\text{Ca}$ and $\text{Re}$, it develops a narrower upper-right end and pinches near the right edge, with its centre displaced below the chamber centreline ($H^* = 0.88$). This is consistent with buoyancy-induced asymmetry favouring edge break-up at lower $\text{Ca/Bo}$. At the lowest viscosity ratios ($\mu_r = 10^{-6}-10^{-5}$), where the droplet offers the least resistance to the host flow, thinning can occur at more than one position along the stretched droplet, producing multiple break-up (Sims 61 and 85). Rotation and retraction occur when the stretched droplet does not pinch off. The lowest orientation angles belong to the rotation regime (Fig.~\ref{fig:fig11}): after maximum elongation at this orientation the droplet keeps rotating, enters the compressional field and folds instead of pinching off. On the other hand, interfacial tension is negligible at these $\text{Ca}$, retraction cannot be a capillary relaxation and must be driven by the host flow. Confinement ratio must act as trigger as retraction is only associated with lower confinement and rotation with higher confinement.
Confinement also changes the strain history and the kinematic path leading to break-up, in addition to the deformed shape. $\gamma_{\max}$ increases with $\text{Ca/Bo}$ as shear overpowers buoyancy, and this dependence is stronger at the higher vertical confinement (Fig.~\ref{fig:fig12}), although the weak trend at $2a/H = 0.12$ is comparable to its scatter, so the difference between the two confinements is indicative.

\subsection{Novelty and Implications for natural enclaves}
Earlier works on droplets have used moderate $\text{Ca}$ and identified a critical capillary number above which a droplet in shear breaks up, and show that this critical value depends on the viscosity ratio and on confinement \citep{amani2019dns, wang2023dynamics}. At these values, surface tension has significant effect on the break up dynamics. However, interfacial tension is very low in magmatic system \citep{edwards2020magma, toramaru2023new}.The simulations here lie far above any such threshold ($\text{Ca}$ from $10^6$ to $8 \times 10^9$), so break-up is not controlled by capillary stability but by how much strain the droplet accumulates. This is why the strain at maximum stretch (on average about $1.1$ to $1.4$, Fig.~\ref{fig:fig12}). We have chosen $\text{Re}$ applicable to very low inertia condition of plutonic system \cite{glazner2014magmatic} which is not common in droplet studies. The present results add buoyancy as a competing control and discusses deformation dependence on $\text{Ca/Bo}$ and confinement.  As a result, simulations produce a sinking regime that neutrally buoyant studies cannot capture, and they show that rotation and retraction occur under controls that differ from the capillary relaxation familiar from low-$\text{Ca}$ studies. The dependence of the break-up mode on viscosity ratio is consistent with the analogue experiments of \citep{hodge2012experimental, hodge2012magmatic}, who found that the break-up of injected mafic dykes depends strongly on the rheological contrast between the magmas, and with the general expectation that a less viscous droplet in a more viscous host deforms readily and ends in pinch-off at its tips \cite{stone1994dynamics, guido2011shear}. 
Our results extends the existing droplet studies to quantify the relationship between orientation, $\text{Ca/Bo}$ and strain at maximum stretching for the two confinement ratios considered here. The lowest orientation angles are associated with the rotation regime, particularly for \(2a/H = 0.20\), where this regime is clearly separated from the other deformation modes.
Together the regimes give a way to read enclave shape in terms of the flow that produced it (Table 4). Buoyancy-dominated descent ($\text{Ca/Bo} < 1$) to the floor of the chamber causes the sagging or flattening of the enclaves along shear direction (Fig.~\ref{fig:fig1}b,d,e). They do not by themselves indicate weak flow or low strain but a local low viscosity host. Strongly fragmented swarms with chains of small fragments point to shear-dominated flow by high viscosity host, whereas two comparable fragments, or a main body with a small satellite, indicate mid or edge break-up at lower $\text{Ca/Bo}$. Moderately elongated and aligned enclaves (Fig.~\ref{fig:fig1}a) may have experienced large accumulated strain, because in the rotation and retraction regimes the droplet does not break up and retracts (Fig.~\ref{fig:fig1}c) or folds (Fig.~\ref{fig:fig1}f) after maximum stretch. Retracted enclaves may have multi-pronged structure (Fig.~\ref{fig:fig1}c) and the aspect ratio of such enclaves can underestimate finite strain. Our results also analyse the claim by \citep{paterson2004significance} that mafic enclaves can't be used as strain markers. Their argument is based on that enclaves, with varied initial shapes and times of formation, undergo large temperature-dependent changes in viscosity contrast. Our simulations considers the hot and fluid stage of enclaves with $\mu_r = 10^{-6}-10^{-3}$, that lie within the “enclave strain” field of \citep{scaillet2000phase}. Because all drops start as identical circles, the variety of outcomes in Figs.~\ref{fig:fig3} -~\ref{fig:fig8} arises from flow regime alone, so shape is non-unique even without differences in initial shape or history. Folded and multi-lobed forms, including “double-pronged” enclaves linked to two successive foliations as described by \citep{paterson2004significance} , can also arise from a single flow in our retraction and folding regimes, so natural examples should be evaluated against both possibilities.

\subsection{Limitations}

The results should be interpreted in the context of the following simplifying assumptions and limitations. The model is two-dimensional, so the out-of-plane stretching and capillary instabilities are absent. The fluids are Newtonian and isothermal with a single density contrast ($400~\text{kg m}^{-3}$), and crystallisation and thermal equilibration, which change the enclave viscosity in natural chambers, are not included. The lateral confinement $2a/L$ was varied through the chamber length, whereas the vertical confinement $2a/H$ changed through the droplet radius, so the vertical confinement is partly combined with droplet size and $\text{Bo}$. The regimes are identified using $\text{Ca/Bo}$, with $H^*$ (Fig.~\ref{fig:fig10}) providing an independent measure of the sinking transition. Finally, natural magma chambers may contain multiple interacting mafic enclaves, whereas the present study considers a single enclave to isolate its deformation and kinematics under the combined effects of buoyancy, confinement, and viscosity contrast. Extending the present framework to multiple interacting enclaves would be a valuable direction for future work, allowing the influence of inter-enclave interactions on deformation, motion, and breakup to be investigated.


\begin{table*}[htbp]
\centering
\footnotesize
\setlength{\tabcolsep}{5pt}
\renewcommand{\arraystretch}{1.2}
\begin{tabular}{|l|l|p{3.0cm}|p{3.0cm}|p{3.0cm}|}
\hline
\rowcolor{gray!20}
\multicolumn{2}{|c|}{\textbf{Regime}} & 
\textbf{Conditions in} \newline \textbf{the simulations} & 
\textbf{Droplet} \newline \textbf{response} & 
\textbf{Expected} \newline \textbf{enclave appearance} \\ \hline
\multicolumn{2}{|l|}{Sinking} & $\text{Ca/Bo} < 1$; all $\mu_r = 10^{-3}$ runs with low $\text{Ca/Bo}$ & Buoyancy dominates; descends toward the base like sagging , sometimes drifting sideways & Rounded to peanut shaped to long flattened enclaves \\ \hline
\multirow{5}{*}{\shortstack[l]{Non-\\Sinking}} & Edge break-up & Mostly $ \text{Ca/Bo} < 5$, limited to $\mu_r = 10^{-4}$ to $10^{-3}$ & Asymmetric thinning; pinch-off near one end & Large main body with a small satellite \\ \cline{2-5}
 & Mid break-up & $\text{Ca/Bo} $ spans from low to high ; main outcome at mid to low $\mu_r = 10^{-6}$ to $10^{-4}$  & Symmetric thinning; pinch-off near the middle of the long axis & Two comparable fragments \\ \cline{2-5}
 & Multi break-up & Mostly $\text{Ca/Bo}> >1$ ; $\mu_r = 10^{-6}$ to $10^{-5}$ & Necking at several positions along the stretched droplet & Chain of fragments \\ \cline{2-5}
 & Rotation & $\text{Ca/Bo} > 1$ but spans over large range ; higher confinement ($2a/H = 0.2$); $\mu_r = 10^{-6}$ to $10^{-4}$ & Stretches, then rotates and folds instead of pinching off & Strongly aligned enclaves, sometimes appear folded\\ \cline{2-5}
 & Retraction & $\text{Ca/Bo} > 1$; mainly lower confinement ($2a/H = 0.12$); $\mu_r = 10^{-6}$ to $10^{-5}$ & Stretches, then shortens toward a compact shape & Compact enclave despite high  accumulated strain \\ \hline
\end{tabular}
\caption{Deformation regimes, the conditions under which they occur in the simulations, and the enclave appearance they would produce. }
\label{tab:table4}
\end{table*}

\section{Conclusions}
\label{sec:conclusions}

We have presented a systematic numerical study of how a dense, low-viscosity magma droplet deforms in a lighter, more viscous host under shear and buoyancy, in the regime of mafic enclaves in granitic plutons ($\mu_r = 10^{-6}-10^{-3}$, $\text{Ca}$ up to $8 \times 10^9$, low $\text{Re}$, $\Delta\rho = 400~\text{kg m}^{-3}$). The $96$ two-dimensional Level-Set simulations resolve six distinct outcomes (sinking, edge, mid and multiple break-up, rotation and retraction) and organise them in terms of $\text{Ca/Bo}$, the viscosity ratio and the confinement. The main results are as follows:

\begin{enumerate}
    \item $\text{Ca/Bo}$ separates gravity-dominated from shear-dominated behaviour. Enclaves with $\text{Ca/Bo} < 1$ sink while retaining much of their shape, reaching depths as low as $H^* \approx 0.2$, whereas those with $\text{Ca/Bo} > 1$ are stretched and reoriented by the host flow. The normalised sinking depth shows that the transition is a range in $\text{Ca/Bo}$, shifted by vertical confinement, and not a single threshold.
    \item The viscosity ratio sets the mode of failure in the shear-dominated regime. Lower $\mu_r$ promotes stretching, break-up is the most common outcome, single pinch-off (edge or mid) occurs across the full $\mu_r$ range, and multiple pinch-off appears only at $\mu_r = 10^{-6}-10^{-5}$, producing chains of fragments.
    \item Confinement decides how the accumulated strain is released. Retraction occurs at lower confinement and rotation at higher confinement, so the two are mutually exclusive. The strain at maximum stretch grows with $\log_{10}(\text{Ca/Bo})$, and this dependence is stronger at $2a/H = 0.20$ (slope $0.31$) than at $2a/H = 0.12$ (slope $0.07$). The orientation of the major axis falls linearly with strain at similar rates for both confinements ($-10.5$ to $-9.5^\circ$ per unit strain).
    \item Together these results turn the diversity of enclave shapes in granitic plutons into a classification that can be read in terms of flow. Rounded, nearly undeformed enclaves can record buoyancy-dominated descent, chains of fragments point to shear-dominated flow at very low viscosity ratio, and moderately elongated enclaves can reflect large accumulated strain, so enclave shape alone is not a direct proxy for finite strain.
\end{enumerate}

The framework rests on a model that is two-dimensional, Newtonian and isothermal. Extending it to three dimensions, to temperature-dependent and crystal-bearing rheology and to a variable density contrast, and testing the regimes against morphometric data from natural enclaves, are natural next steps.

\section*{Acknowledgement}
\label{sec:ack}
\noindent

UD acknowledges CSIR-India for ASPIRE research grant no. 25WS(0009)/2023-24/EMR-II/ASPIRE. PS acknowledges for the financial support through ASPIRE research grant no.  25WS(0009)/2023-24/EMR-II/ASPIRE.
\newpage

\appendix
\section{Complete List of input and output parameters}
\label{sec:appendix1}


\scriptsize
\setlength{\tabcolsep}{3pt}
\renewcommand{\arraystretch}{1.1}
\begin{longtable}{||p{1.8cm}||p{1.6cm}||p{1.2cm}||p{1.6cm}||p{1.8cm}||p{1.8cm}||p{1.8cm}||}
\hline
\hline
\rowcolor{gray!20}
\textbf{Simulation} & \textbf{$\mu_r$} & \textbf{$2a/H$} & \textbf{$2a/L$} & \textbf{$\text{Ca}$} & \textbf{$\text{Bo}$} & \textbf{$\text{Re}$} \\ \hline
\endfirsthead
\hline
\rowcolor{gray!20}
\textbf{Simulation} & \textbf{$\mu_r$} & \textbf{$2a/H$} & \textbf{$2a/L$} & \textbf{$\text{Ca}$} & \textbf{$\text{Bo}$} & \textbf{$\text{Re}$} \\ \hline
\endhead
\hline \multicolumn{7}{r}{\textit{Continued on next page}} \\ \hline
\endfoot
\hline
\endlastfoot
SIM 1 & $1.00 \times 10^{-3}$ & $0.12$ & $7.50 \times 10^{-2}$ & $1.00 \times 10^6$ & $3.53 \times 10^6$ & $1.15 \times 10^{-3}$ \\ \hline
SIM 2 & $1.00 \times 10^{-3}$ & $0.12$ & $7.50 \times 10^{-2}$ & $3.00 \times 10^6$ & $3.53 \times 10^6$ & $3.45 \times 10^{-3}$ \\ \hline
SIM 3 & $1.00 \times 10^{-3}$ & $0.12$ & $7.50 \times 10^{-2}$ & $5.00 \times 10^6$ & $3.53 \times 10^6$ & $5.75 \times 10^{-3}$ \\ \hline
SIM 4 & $1.00 \times 10^{-3}$ & $0.12$ & $7.50 \times 10^{-2}$ & $8.00 \times 10^6$ & $3.53 \times 10^6$ & $9.20 \times 10^{-3}$ \\ \hline
SIM 5 & $1.00 \times 10^{-3}$ & $0.2$ & $1.25 \times 10^{-1}$ & $1.00 \times 10^6$ & $9.81 \times 10^6$ & $1.15 \times 10^{-3}$ \\ \hline
SIM 6 & $1.00 \times 10^{-3}$ & $0.2$ & $1.25 \times 10^{-1}$ & $3.00 \times 10^6$ & $9.81 \times 10^6$ & $3.45 \times 10^{-3}$ \\ \hline
SIM 7 & $1.00 \times 10^{-3}$ & $0.2$ & $1.25 \times 10^{-1}$ & $5.00 \times 10^6$ & $9.81 \times 10^6$ & $5.75 \times 10^{-3}$ \\ \hline
SIM 8 & $1.00 \times 10^{-3}$ & $0.2$ & $1.25 \times 10^{-1}$ & $8.00 \times 10^6$ & $9.81 \times 10^6$ & $9.20 \times 10^{-3}$ \\ \hline
SIM 9 & $1.00 \times 10^{-3}$ & $0.12$ & $6.00 \times 10^{-2}$ & $1.00 \times 10^6$ & $3.53 \times 10^6$ & $1.15 \times 10^{-3}$ \\ \hline
SIM 10 & $1.00 \times 10^{-3}$ & $0.12$ & $6.00 \times 10^{-2}$ & $3.00 \times 10^6$ & $3.53 \times 10^6$ & $3.45 \times 10^{-3}$ \\ \hline
SIM 11 & $1.00 \times 10^{-3}$ & $0.12$ & $6.00 \times 10^{-2}$ & $5.00 \times 10^6$ & $3.53 \times 10^6$ & $5.75 \times 10^{-3}$ \\ \hline
SIM 12 & $1.00 \times 10^{-3}$ & $0.12$ & $6.00 \times 10^{-2}$ & $8.00 \times 10^6$ & $3.53 \times 10^6$ & $9.20 \times 10^{-3}$ \\ \hline
SIM 13 & $1.00 \times 10^{-3}$ & $0.2$ & $1.00 \times 10^{-1}$ & $1.00 \times 10^6$ & $9.81 \times 10^6$ & $1.15 \times 10^{-3}$ \\ \hline
SIM 14 & $1.00 \times 10^{-3}$ & $0.2$ & $1.00 \times 10^{-1}$ & $3.00 \times 10^6$ & $9.81 \times 10^6$ & $3.45 \times 10^{-3}$ \\ \hline
SIM 15 & $1.00 \times 10^{-3}$ & $0.2$ & $1.00 \times 10^{-1}$ & $5.00 \times 10^6$ & $9.81 \times 10^6$ & $5.75 \times 10^{-3}$ \\ \hline
SIM 16 & $1.00 \times 10^{-3}$ & $0.2$ & $1.00 \times 10^{-1}$ & $8.00 \times 10^6$ & $9.81 \times 10^6$ & $9.20 \times 10^{-3}$ \\ \hline
SIM 17 & $1.00 \times 10^{-3}$ & $0.12$ & $4.00 \times 10^{-2}$ & $1.00 \times 10^6$ & $3.53 \times 10^6$ & $1.15 \times 10^{-3}$ \\ \hline
SIM 18 & $1.00 \times 10^{-3}$ & $0.12$ & $4.00 \times 10^{-2}$ & $3.00 \times 10^6$ & $3.53 \times 10^6$ & $3.45 \times 10^{-3}$ \\ \hline
SIM 19 & $1.00 \times 10^{-3}$ & $0.12$ & $4.00 \times 10^{-2}$ & $5.00 \times 10^6$ & $3.53 \times 10^6$ & $5.75 \times 10^{-3}$ \\ \hline
SIM 20 & $1.00 \times 10^{-3}$ & $0.12$ & $4.00 \times 10^{-2}$ & $8.00 \times 10^6$ & $3.53 \times 10^6$ & $9.20 \times 10^{-3}$ \\ \hline
SIM 21 & $1.00 \times 10^{-3}$ & $0.2$ & $6.67 \times 10^{-2}$ & $1.00 \times 10^6$ & $9.81 \times 10^6$ & $1.15 \times 10^{-3}$ \\ \hline
SIM 22 & $1.00 \times 10^{-3}$ & $0.2$ & $6.67 \times 10^{-2}$ & $3.00 \times 10^6$ & $9.81 \times 10^6$ & $3.45 \times 10^{-3}$ \\ \hline
SIM 23 & $1.00 \times 10^{-3}$ & $0.2$ & $6.67 \times 10^{-2}$ & $5.00 \times 10^6$ & $9.81 \times 10^6$ & $5.75 \times 10^{-3}$ \\ \hline
SIM 24 & $1.00 \times 10^{-3}$ & $0.2$ & $6.67 \times 10^{-2}$ & $8.00 \times 10^6$ & $9.81 \times 10^6$ & $9.20 \times 10^{-3}$ \\ \hline
SIM 25 & $1.00 \times 10^{-4}$ & $0.12$ & $7.50 \times 10^{-2}$ & $1.00 \times 10^7$ & $3.53 \times 10^6$ & $1.15 \times 10^{-4}$ \\ \hline
SIM 26 & $1.00 \times 10^{-4}$ & $0.12$ & $7.50 \times 10^{-2}$ & $3.00 \times 10^7$ & $3.53 \times 10^6$ & $3.45 \times 10^{-4}$ \\ \hline
SIM 27 & $1.00 \times 10^{-4}$ & $0.12$ & $7.50 \times 10^{-2}$ & $5.00 \times 10^7$ & $3.53 \times 10^6$ & $5.75 \times 10^{-4}$ \\ \hline
SIM 28 & $1.00 \times 10^{-4}$ & $0.12$ & $7.50 \times 10^{-2}$ & $8.00 \times 10^7$ & $3.53 \times 10^6$ & $9.20 \times 10^{-4}$ \\ \hline
SIM 29 & $1.00 \times 10^{-4}$ & $0.2$ & $1.25 \times 10^{-1}$ & $1.00 \times 10^7$ & $9.81 \times 10^6$ & $1.15 \times 10^{-4}$ \\ \hline
SIM 30 & $1.00 \times 10^{-4}$ & $0.2$ & $1.25 \times 10^{-1}$ & $3.00 \times 10^7$ & $9.81 \times 10^6$ & $3.45 \times 10^{-4}$ \\ \hline
SIM 31 & $1.00 \times 10^{-4}$ & $0.2$ & $1.25 \times 10^{-1}$ & $5.00 \times 10^7$ & $9.81 \times 10^6$ & $5.75 \times 10^{-4}$ \\ \hline
SIM 32 & $1.00 \times 10^{-4}$ & $0.2$ & $1.25 \times 10^{-1}$ & $8.00 \times 10^7$ & $9.81 \times 10^6$ & $9.20 \times 10^{-4}$ \\ \hline
SIM 33 & $1.00 \times 10^{-4}$ & $0.12$ & $6.00 \times 10^{-2}$ & $1.00 \times 10^7$ & $3.53 \times 10^6$ & $1.15 \times 10^{-4}$ \\ \hline
SIM 34 & $1.00 \times 10^{-4}$ & $0.12$ & $6.00 \times 10^{-2}$ & $3.00 \times 10^7$ & $3.53 \times 10^6$ & $3.45 \times 10^{-4}$ \\ \hline
SIM 35 & $1.00 \times 10^{-4}$ & $0.12$ & $6.00 \times 10^{-2}$ & $5.00 \times 10^7$ & $3.53 \times 10^6$ & $5.75 \times 10^{-4}$ \\ \hline
SIM 36 & $1.00 \times 10^{-4}$ & $0.12$ & $6.00 \times 10^{-2}$ & $8.00 \times 10^7$ & $3.53 \times 10^6$ & $9.20 \times 10^{-4}$ \\ \hline
SIM 37 & $1.00 \times 10^{-4}$ & $0.2$ & $1.00 \times 10^{-1}$ & $1.00 \times 10^7$ & $9.81 \times 10^6$ & $1.15 \times 10^{-4}$ \\ \hline
SIM 38 & $1.00 \times 10^{-4}$ & $0.2$ & $1.00 \times 10^{-1}$ & $3.00 \times 10^7$ & $9.81 \times 10^6$ & $3.45 \times 10^{-4}$ \\ \hline
SIM 39 & $1.00 \times 10^{-4}$ & $0.2$ & $1.00 \times 10^{-1}$ & $5.00 \times 10^7$ & $9.81 \times 10^6$ & $5.75 \times 10^{-4}$ \\ \hline
SIM 40 & $1.00 \times 10^{-4}$ & $0.2$ & $1.00 \times 10^{-1}$ & $8.00 \times 10^7$ & $9.81 \times 10^6$ & $9.20 \times 10^{-4}$ \\ \hline
SIM 41 & $1.00 \times 10^{-4}$ & $0.12$ & $4.00 \times 10^{-2}$ & $1.00 \times 10^7$ & $3.53 \times 10^6$ & $1.15 \times 10^{-4}$ \\ \hline
SIM 42 & $1.00 \times 10^{-4}$ & $0.12$ & $4.00 \times 10^{-2}$ & $3.00 \times 10^7$ & $3.53 \times 10^6$ & $3.45 \times 10^{-4}$ \\ \hline
SIM 43 & $1.00 \times 10^{-4}$ & $0.12$ & $4.00 \times 10^{-2}$ & $5.00 \times 10^7$ & $3.53 \times 10^6$ & $5.75 \times 10^{-4}$ \\ \hline
SIM 44 & $1.00 \times 10^{-4}$ & $0.12$ & $4.00 \times 10^{-2}$ & $8.00 \times 10^7$ & $3.53 \times 10^6$ & $9.20 \times 10^{-4}$ \\ \hline
SIM 45 & $1.00 \times 10^{-4}$ & $0.2$ & $6.67 \times 10^{-2}$ & $1.00 \times 10^7$ & $9.81 \times 10^6$ & $1.15 \times 10^{-4}$ \\ \hline
SIM 46 & $1.00 \times 10^{-4}$ & $0.2$ & $6.67 \times 10^{-2}$ & $3.00 \times 10^7$ & $9.81 \times 10^6$ & $3.45 \times 10^{-4}$ \\ \hline
SIM 47 & $1.00 \times 10^{-4}$ & $0.2$ & $6.67 \times 10^{-2}$ & $5.00 \times 10^7$ & $9.81 \times 10^6$ & $5.75 \times 10^{-4}$ \\ \hline
SIM 48 & $1.00 \times 10^{-4}$ & $0.2$ & $6.67 \times 10^{-2}$ & $8.00 \times 10^7$ & $9.81 \times 10^6$ & $9.20 \times 10^{-4}$ \\ \hline
SIM 49 & $1.00 \times 10^{-5}$ & $0.12$ & $7.50 \times 10^{-2}$ & $1.00 \times 10^8$ & $3.53 \times 10^6$ & $1.15 \times 10^{-5}$ \\ \hline
SIM 50 & $1.00 \times 10^{-5}$ & $0.12$ & $7.50 \times 10^{-2}$ & $3.00 \times 10^8$ & $3.53 \times 10^6$ & $3.45 \times 10^{-5}$ \\ \hline
SIM 51 & $1.00 \times 10^{-5}$ & $0.12$ & $7.50 \times 10^{-2}$ & $5.00 \times 10^8$ & $3.53 \times 10^6$ & $5.75 \times 10^{-5}$ \\ \hline
SIM 52 & $1.00 \times 10^{-5}$ & $0.12$ & $7.50 \times 10^{-2}$ & $8.00 \times 10^8$ & $3.53 \times 10^6$ & $9.20 \times 10^{-5}$ \\ \hline
SIM 53 & $1.00 \times 10^{-5}$ & $0.2$ & $1.25 \times 10^{-1}$ & $1.00 \times 10^8$ & $9.81 \times 10^6$ & $1.15 \times 10^{-5}$ \\ \hline
SIM 54 & $1.00 \times 10^{-5}$ & $0.2$ & $1.25 \times 10^{-1}$ & $3.00 \times 10^8$ & $9.81 \times 10^6$ & $3.45 \times 10^{-5}$ \\ \hline
SIM 55 & $1.00 \times 10^{-5}$ & $0.2$ & $1.25 \times 10^{-1}$ & $5.00 \times 10^8$ & $9.81 \times 10^6$ & $5.75 \times 10^{-5}$ \\ \hline
SIM 56 & $1.00 \times 10^{-5}$ & $0.2$ & $1.25 \times 10^{-1}$ & $8.00 \times 10^8$ & $9.81 \times 10^6$ & $9.20 \times 10^{-5}$ \\ \hline
SIM 57 & $1.00 \times 10^{-5}$ & $0.12$ & $6.00 \times 10^{-2}$ & $1.00 \times 10^8$ & $3.53 \times 10^6$ & $1.15 \times 10^{-5}$ \\ \hline
SIM 58 & $1.00 \times 10^{-5}$ & $0.12$ & $6.00 \times 10^{-2}$ & $3.00 \times 10^8$ & $3.53 \times 10^6$ & $3.45 \times 10^{-5}$ \\ \hline
SIM 59 & $1.00 \times 10^{-5}$ & $0.12$ & $6.00 \times 10^{-2}$ & $5.00 \times 10^8$ & $3.53 \times 10^6$ & $5.75 \times 10^{-5}$ \\ \hline
SIM 60 & $1.00 \times 10^{-5}$ & $0.12$ & $6.00 \times 10^{-2}$ & $8.00 \times 10^8$ & $3.53 \times 10^6$ & $9.20 \times 10^{-5}$ \\ \hline
SIM 61 & $1.00 \times 10^{-5}$ & $0.2$ & $1.00 \times 10^{-1}$ & $1.00 \times 10^8$ & $9.81 \times 10^6$ & $1.15 \times 10^{-5}$ \\ \hline
SIM 62 & $1.00 \times 10^{-5}$ & $0.2$ & $1.00 \times 10^{-1}$ & $3.00 \times 10^8$ & $9.81 \times 10^6$ & $3.45 \times 10^{-5}$ \\ \hline
SIM 63 & $1.00 \times 10^{-5}$ & $0.2$ & $1.00 \times 10^{-1}$ & $5.00 \times 10^8$ & $9.81 \times 10^6$ & $5.75 \times 10^{-5}$ \\ \hline
SIM 64 & $1.00 \times 10^{-5}$ & $0.2$ & $1.00 \times 10^{-1}$ & $8.00 \times 10^8$ & $9.81 \times 10^6$ & $9.20 \times 10^{-5}$ \\ \hline
SIM 65 & $1.00 \times 10^{-5}$ & $0.12$ & $4.00 \times 10^{-2}$ & $1.00 \times 10^8$ & $3.53 \times 10^6$ & $1.15 \times 10^{-5}$ \\ \hline
SIM 66 & $1.00 \times 10^{-5}$ & $0.12$ & $4.00 \times 10^{-2}$ & $3.00 \times 10^8$ & $3.53 \times 10^6$ & $3.45 \times 10^{-5}$ \\ \hline
SIM 67 & $1.00 \times 10^{-5}$ & $0.12$ & $4.00 \times 10^{-2}$ & $5.00 \times 10^8$ & $3.53 \times 10^6$ & $5.75 \times 10^{-5}$ \\ \hline
SIM 68 & $1.00 \times 10^{-5}$ & $0.12$ & $4.00 \times 10^{-2}$ & $8.00 \times 10^8$ & $3.53 \times 10^6$ & $9.20 \times 10^{-5}$ \\ \hline
SIM 69 & $1.00 \times 10^{-5}$ & $0.2$ & $6.67 \times 10^{-2}$ & $1.00 \times 10^8$ & $9.81 \times 10^6$ & $1.15 \times 10^{-5}$ \\ \hline
SIM 70 & $1.00 \times 10^{-5}$ & $0.2$ & $6.67 \times 10^{-2}$ & $3.00 \times 10^8$ & $9.81 \times 10^6$ & $3.45 \times 10^{-5}$ \\ \hline
SIM 71 & $1.00 \times 10^{-5}$ & $0.2$ & $6.67 \times 10^{-2}$ & $5.00 \times 10^8$ & $9.81 \times 10^6$ & $5.75 \times 10^{-5}$ \\ \hline
SIM 72 & $1.00 \times 10^{-5}$ & $0.2$ & $6.67 \times 10^{-2}$ & $8.00 \times 10^8$ & $9.81 \times 10^6$ & $9.20 \times 10^{-5}$ \\ \hline
SIM 73 & $1.00 \times 10^{-6}$ & $0.12$ & $7.50 \times 10^{-2}$ & $1.00 \times 10^9$ & $3.53 \times 10^6$ & $1.15 \times 10^{-6}$ \\ \hline
SIM 74 & $1.00 \times 10^{-6}$ & $0.12$ & $7.50 \times 10^{-2}$ & $3.00 \times 10^9$ & $3.53 \times 10^6$ & $3.45 \times 10^{-6}$ \\ \hline
SIM 75 & $1.00 \times 10^{-6}$ & $0.12$ & $7.50 \times 10^{-2}$ & $5.00 \times 10^9$ & $3.53 \times 10^6$ & $5.75 \times 10^{-6}$ \\ \hline
SIM 76 & $1.00 \times 10^{-6}$ & $0.12$ & $7.50 \times 10^{-2}$ & $8.00 \times 10^9$ & $3.53 \times 10^6$ & $9.20 \times 10^{-6}$ \\ \hline
SIM 77 & $1.00 \times 10^{-6}$ & $0.2$ & $1.25 \times 10^{-1}$ & $1.00 \times 10^9$ & $9.81 \times 10^6$ & $1.15 \times 10^{-6}$ \\ \hline
SIM 78 & $1.00 \times 10^{-6}$ & $0.2$ & $1.25 \times 10^{-1}$ & $3.00 \times 10^9$ & $9.81 \times 10^6$ & $3.45 \times 10^{-6}$ \\ \hline
SIM 79 & $1.00 \times 10^{-6}$ & $0.2$ & $1.25 \times 10^{-1}$ & $5.00 \times 10^9$ & $9.81 \times 10^6$ & $5.75 \times 10^{-6}$ \\ \hline
SIM 80 & $1.00 \times 10^{-6}$ & $0.2$ & $1.25 \times 10^{-1}$ & $8.00 \times 10^9$ & $9.81 \times 10^6$ & $9.20 \times 10^{-6}$ \\ \hline
SIM 81 & $1.00 \times 10^{-6}$ & $0.12$ & $6.00 \times 10^{-2}$ & $1.00 \times 10^9$ & $3.53 \times 10^6$ & $1.15 \times 10^{-6}$ \\ \hline
SIM 82 & $1.00 \times 10^{-6}$ & $0.12$ & $6.00 \times 10^{-2}$ & $3.00 \times 10^9$ & $3.53 \times 10^6$ & $3.45 \times 10^{-6}$ \\ \hline
SIM 83 & $1.00 \times 10^{-6}$ & $0.12$ & $6.00 \times 10^{-2}$ & $5.00 \times 10^9$ & $3.53 \times 10^6$ & $5.75 \times 10^{-6}$ \\ \hline
SIM 84 & $1.00 \times 10^{-6}$ & $0.12$ & $6.00 \times 10^{-2}$ & $8.00 \times 10^9$ & $3.53 \times 10^6$ & $9.20 \times 10^{-6}$ \\ \hline
SIM 85 & $1.00 \times 10^{-6}$ & $0.2$ & $1.00 \times 10^{-1}$ & $1.00 \times 10^9$ & $9.81 \times 10^6$ & $1.15 \times 10^{-6}$ \\ \hline
SIM 86 & $1.00 \times 10^{-6}$ & $0.2$ & $1.00 \times 10^{-1}$ & $3.00 \times 10^9$ & $9.81 \times 10^6$ & $3.45 \times 10^{-6}$ \\ \hline
SIM 87 & $1.00 \times 10^{-6}$ & $0.2$ & $1.00 \times 10^{-1}$ & $5.00 \times 10^9$ & $9.81 \times 10^6$ & $5.75 \times 10^{-6}$ \\ \hline
SIM 88 & $1.00 \times 10^{-6}$ & $0.2$ & $1.00 \times 10^{-1}$ & $8.00 \times 10^9$ & $9.81 \times 10^6$ & $9.20 \times 10^{-6}$ \\ \hline
SIM 89 & $1.00 \times 10^{-6}$ & $0.12$ & $4.00 \times 10^{-2}$ & $1.00 \times 10^9$ & $3.53 \times 10^6$ & $1.15 \times 10^{-6}$ \\ \hline
SIM 90 & $1.00 \times 10^{-6}$ & $0.12$ & $4.00 \times 10^{-2}$ & $3.00 \times 10^9$ & $3.53 \times 10^6$ & $3.45 \times 10^{-6}$ \\ \hline
SIM 91 & $1.00 \times 10^{-6}$ & $0.12$ & $4.00 \times 10^{-2}$ & $5.00 \times 10^9$ & $3.53 \times 10^6$ & $5.75 \times 10^{-6}$ \\ \hline
SIM 92 & $1.00 \times 10^{-6}$ & $0.12$ & $4.00 \times 10^{-2}$ & $8.00 \times 10^9$ & $3.53 \times 10^6$ & $9.20 \times 10^{-6}$ \\ \hline
SIM 93 & $1.00 \times 10^{-6}$ & $0.2$ & $6.67 \times 10^{-2}$ & $1.00 \times 10^9$ & $9.81 \times 10^6$ & $1.15 \times 10^{-6}$ \\ \hline
SIM 94 & $1.00 \times 10^{-6}$ & $0.2$ & $6.67 \times 10^{-2}$ & $3.00 \times 10^9$ & $9.81 \times 10^6$ & $3.45 \times 10^{-6}$ \\ \hline
SIM 95 & $1.00 \times 10^{-6}$ & $0.2$ & $6.67 \times 10^{-2}$ & $5.00 \times 10^9$ & $9.81 \times 10^6$ & $5.75 \times 10^{-6}$ \\ \hline
SIM 96 & $1.00 \times 10^{-6}$ & $0.2$ & $6.67 \times 10^{-2}$ & $8.00 \times 10^9$ & $9.81 \times 10^6$ & $9.20 \times 10^{-6}$ \\ \hline
\caption{Complete simulation input parameters for runs 1 to 96.}
\label{tab:sim_inputs}
\end{longtable}

\scriptsize
\setlength{\tabcolsep}{4pt}
\renewcommand{\arraystretch}{1.1}
\begin{longtable}{||p{2.4cm}||p{2.4cm}||p{2.4cm}||p{2.4cm}||p{2.4cm}||}
\hline
\hline
\rowcolor{gray!20}
\textbf{Simulation} & \textbf{$D$} & \textbf{$\delta$} & \textbf{$\theta$} & \textbf{$\gamma$} \\ \hline
\endfirsthead
\hline
\rowcolor{gray!20}
\textbf{Simulation} & \textbf{$D$} & \textbf{$\delta$} & \textbf{$\theta$} & \textbf{$\gamma$} \\ \hline
\endhead
\hline \multicolumn{5}{r}{\textit{Continued on next page}} \\ \hline
\endfoot
\hline
\endlastfoot
SIM 1 & $0.39$ & $1.39$ & $9.23$ & $5.89 \times 10^{-1}$ \\ \hline
SIM 2 & $0.45$ & $1.44$ & $26.83$ & $1.89 \times 10^{0}$ \\ \hline
SIM 3 & $0.52$ & $1.50$ & $32.71$ & $8.42 \times 10^{-1}$ \\ \hline
SIM 4 & $0.63$ & $1.68$ & $27.87$ & $8.83 \times 10^{-1}$ \\ \hline
SIM 5 & $0.63$ & $1.63$ & $3.54$ & $1.18 \times 10^{0}$ \\ \hline
SIM 6 & $0.70$ & $1.71$ & $18.48$ & $8.84 \times 10^{-1}$ \\ \hline
SIM 7 & $0.71$ & $1.72$ & $29.07$ & $9.47 \times 10^{-1}$ \\ \hline
SIM 8 & $0.75$ & $1.78$ & $33.89$ & $1.10 \times 10^{0}$ \\ \hline
SIM 9 & $0.50$ & $1.51$ & $7.67$ & $6.62 \times 10^{-1}$ \\ \hline
SIM 10 & $0.54$ & $1.56$ & $31.52$ & $7.45 \times 10^{-1}$ \\ \hline
SIM 11 & $0.57$ & $1.62$ & $34.32$ & $9.66 \times 10^{-1}$ \\ \hline
SIM 12 & $0.58$ & $1.76$ & $34.95$ & $9.93 \times 10^{-1}$ \\ \hline
SIM 13 & $0.74$ & $2.16$ & $2.54$ & $4.97 \times 10^{-1}$ \\ \hline
SIM 14 & $0.76$ & $2.21$ & $28.22$ & $1.24 \times 10^{0}$ \\ \hline
SIM 15 & $0.79$ & $2.34$ & $27.26$ & $1.24 \times 10^{0}$ \\ \hline
SIM 16 & $0.58$ & $1.63$ & $33.52$ & $6.62 \times 10^{-1}$ \\ \hline
SIM 17 & $0.51$ & $1.51$ & $6.34$ & $7.17 \times 10^{-1}$ \\ \hline
SIM 18 & $0.64$ & $1.71$ & $14.80$ & $2.07 \times 10^{0}$ \\ \hline
SIM 19 & $0.65$ & $1.81$ & $33.20$ & $9.66 \times 10^{-1}$ \\ \hline
SIM 20 & $0.67$ & $1.89$ & $35.57$ & $8.83 \times 10^{-1}$ \\ \hline
SIM 21 & $0.77$ & $2.28$ & $1.84$ & $5.52 \times 10^{-1}$ \\ \hline
SIM 22 & $0.67$ & $1.93$ & $13.53$ & $9.10 \times 10^{-1}$ \\ \hline
SIM 23 & $0.66$ & $1.78$ & $24.58$ & $8.28 \times 10^{-1}$ \\ \hline
SIM 24 & $0.67$ & $1.91$ & $32.46$ & $8.83 \times 10^{-1}$ \\ \hline
SIM 25 & $0.27$ & $1.27$ & $37.85$ & $1.18 \times 10^{0}$ \\ \hline
SIM 26 & $0.51$ & $1.58$ & $31.17$ & $1.42 \times 10^{0}$ \\ \hline
SIM 27 & $0.57$ & $1.70$ & $31.13$ & $1.26 \times 10^{0}$ \\ \hline
SIM 28 & $0.59$ & $1.80$ & $26.24$ & $1.52 \times 10^{0}$ \\ \hline
SIM 29 & $0.74$ & $0.86$ & $34.07$ & $6.74 \times 10^{-1}$ \\ \hline
SIM 30 & $0.84$ & $0.95$ & $30.87$ & $9.47 \times 10^{-1}$ \\ \hline
SIM 31 & $0.87$ & $2.30$ & $12.34$ & $1.58 \times 10^{0}$ \\ \hline
SIM 32 & $0.88$ & $2.47$ & $-0.39$ & $1.66 \times 10^{0}$ \\ \hline
SIM 33 & $0.34$ & $1.34$ & $34.66$ & $1.43 \times 10^{0}$ \\ \hline
SIM 34 & $0.41$ & $1.45$ & $39.77$ & $4.14 \times 10^{-1}$ \\ \hline
SIM 35 & $0.42$ & $1.49$ & $37.54$ & $6.21 \times 10^{-1}$ \\ \hline
SIM 36 & $0.45$ & $1.55$ & $37.45$ & $6.62 \times 10^{-1}$ \\ \hline
SIM 37 & $0.76$ & $1.48$ & $38.75$ & $5.79 \times 10^{-1}$ \\ \hline
SIM 38 & $0.84$ & $2.00$ & $29.42$ & $1.12 \times 10^{0}$ \\ \hline
SIM 39 & $0.85$ & $2.14$ & $28.84$ & $1.24 \times 10^{0}$ \\ \hline
SIM 40 & $0.91$ & $2.26$ & $28.00$ & $1.32 \times 10^{0}$ \\ \hline
SIM 41 & $0.42$ & $1.45$ & $35.89$ & $1.24 \times 10^{0}$ \\ \hline
SIM 42 & $0.61$ & $1.76$ & $34.03$ & $1.24 \times 10^{0}$ \\ \hline
SIM 43 & $0.62$ & $1.83$ & $32.32$ & $1.03 \times 10^{0}$ \\ \hline
SIM 44 & $0.71$ & $1.96$ & $33.41$ & $1.10 \times 10^{0}$ \\ \hline
SIM 45 & $0.42$ & $1.53$ & $36.87$ & $6.21 \times 10^{-1}$ \\ \hline
SIM 46 & $0.62$ & $1.94$ & $32.07$ & $9.47 \times 10^{-1}$ \\ \hline
SIM 47 & $0.64$ & $2.05$ & $32.24$ & $1.03 \times 10^{0}$ \\ \hline
SIM 48 & $0.80$ & $2.71$ & $26.39$ & $1.66 \times 10^{0}$ \\ \hline
SIM 49 & $0.13$ & $2.34$ & $40.08$ & $1.01 \times 10^{0}$ \\ \hline
SIM 50 & $0.57$ & $1.60$ & $23.33$ & $1.61 \times 10^{0}$ \\ \hline
SIM 51 & $0.60$ & $1.72$ & $26.12$ & $1.42 \times 10^{0}$ \\ \hline
SIM 52 & $0.69$ & $1.91$ & $22.39$ & $1.54 \times 10^{0}$ \\ \hline
SIM 53 & $0.57$ & $1.80$ & $25.60$ & $1.47 \times 10^{0}$ \\ \hline
SIM 54 & $0.62$ & $1.91$ & $0.44$ & $1.45 \times 10^{0}$ \\ \hline
SIM 55 & $0.64$ & $1.99$ & $3.60$ & $1.58 \times 10^{0}$ \\ \hline
SIM 56 & $0.66$ & $2.06$ & $2.36$ & $1.66 \times 10^{0}$ \\ \hline
SIM 57 & $0.38$ & $1.33$ & $40.36$ & $1.82 \times 10^{0}$ \\ \hline
SIM 58 & $0.65$ & $1.46$ & $38.61$ & $6.21 \times 10^{-1}$ \\ \hline
SIM 59 & $0.67$ & $1.53$ & $37.25$ & $6.90 \times 10^{-1}$ \\ \hline
SIM 60 & $0.70$ & $1.87$ & $32.72$ & $5.52 \times 10^{-1}$ \\ \hline
SIM 61 & $0.63$ & $1.96$ & $29.05$ & $1.24 \times 10^{0}$ \\ \hline
SIM 62 & $0.67$ & $2.09$ & $29.43$ & $1.24 \times 10^{0}$ \\ \hline
SIM 63 & $0.71$ & $2.24$ & $27.18$ & $1.38 \times 10^{0}$ \\ \hline
SIM 64 & $0.77$ & $2.53$ & $23.70$ & $1.66 \times 10^{0}$ \\ \hline
SIM 65 & $0.53$ & $1.51$ & $41.87$ & $7.59 \times 10^{-1}$ \\ \hline
SIM 66 & $0.65$ & $1.83$ & $32.85$ & $1.24 \times 10^{0}$ \\ \hline
SIM 67 & $0.63$ & $1.79$ & $30.29$ & $1.03 \times 10^{0}$ \\ \hline
SIM 68 & $0.71$ & $1.96$ & $33.75$ & $1.10 \times 10^{0}$ \\ \hline
SIM 69 & $0.83$ & $2.40$ & $27.27$ & $1.72 \times 10^{0}$ \\ \hline
SIM 70 & $0.87$ & $2.75$ & $24.77$ & $1.86 \times 10^{0}$ \\ \hline
SIM 71 & $0.85$ & $2.70$ & $25.70$ & $1.72 \times 10^{0}$ \\ \hline
SIM 72 & $0.86$ & $3.11$ & $23.14$ & $2.02 \times 10^{0}$ \\ \hline
SIM 73 & $0.25$ & $1.22$ & $32.39$ & $2.59 \times 10^{0}$ \\ \hline
SIM 74 & $0.44$ & $1.52$ & $34.67$ & $1.01 \times 10^{0}$ \\ \hline
SIM 75 & $0.55$ & $1.67$ & $30.29$ & $1.26 \times 10^{0}$ \\ \hline
SIM 76 & $0.61$ & $1.79$ & $24.56$ & $1.54 \times 10^{0}$ \\ \hline
SIM 77 & $0.63$ & $1.83$ & $22.71$ & $1.68 \times 10^{0}$ \\ \hline
SIM 78 & $0.72$ & $2.12$ & $1.11$ & $2.02 \times 10^{0}$ \\ \hline
SIM 79 & $0.76$ & $2.30$ & $1.55$ & $2.53 \times 10^{0}$ \\ \hline
SIM 80 & $0.77$ & $2.40$ & $2.69$ & $1.99 \times 10^{0}$ \\ \hline
SIM 81 & $0.41$ & $1.33$ & $40.40$ & $1.60 \times 10^{0}$ \\ \hline
SIM 82 & $0.58$ & $1.61$ & $33.33$ & $9.93 \times 10^{-1}$ \\ \hline
SIM 83 & $0.61$ & $1.71$ & $33.10$ & $9.66 \times 10^{-1}$ \\ \hline
SIM 84 & $0.63$ & $1.79$ & $34.02$ & $9.93 \times 10^{-1}$ \\ \hline
SIM 85 & $0.75$ & $2.31$ & $26.05$ & $1.77 \times 10^{0}$ \\ \hline
SIM 86 & $0.77$ & $2.37$ & $25.51$ & $1.61 \times 10^{0}$ \\ \hline
SIM 87 & $0.78$ & $2.46$ & $24.59$ & $1.66 \times 10^{0}$ \\ \hline
SIM 88 & $0.81$ & $2.53$ & $23.68$ & $1.66 \times 10^{0}$ \\ \hline
SIM 89 & $0.43$ & $1.48$ & $42.54$ & $1.77 \times 10^{0}$ \\ \hline
SIM 90 & $0.61$ & $1.75$ & $31.42$ & $1.66 \times 10^{0}$ \\ \hline
SIM 91 & $0.65$ & $1.91$ & $35.93$ & $1.10 \times 10^{0}$ \\ \hline
SIM 92 & $0.68$ & $1.93$ & $34.04$ & $1.10 \times 10^{0}$ \\ \hline
SIM 93 & $0.74$ & $2.30$ & $26.57$ & $1.66 \times 10^{0}$ \\ \hline
SIM 94 & $0.80$ & $2.58$ & $24.98$ & $1.66 \times 10^{0}$ \\ \hline
SIM 95 & $0.82$ & $2.82$ & $25.58$ & $1.79 \times 10^{0}$ \\ \hline
SIM 96 & $0.83$ & $2.82$ & $25.95$ & $1.77 \times 10^{0}$ \\ \hline
\caption{Complete simulation output parameters for Simulation-1 to 96.}
\label{tab:sim_outputs}
\end{longtable}

\bibliography{refs}

\providecommand{\href}[2]{#2}\begingroup\raggedright\begin{thebibliography}{10}

\bibitem{barbarin1992genesis}
B.~Barbarin and J.~Didier, \emph{Genesis and evolution of mafic microgranular
  enclaves through various types of interaction between coexisting felsic and
  mafic magmas}, {\emph{Earth and Environmental Science Transactions of the
  Royal Society of Edinburgh} {\bf 83} (1992) 145--153}.

\bibitem{barbarin2005mafic}
B.~Barbarin, \emph{Mafic magmatic enclaves and mafic rocks associated with some
  granitoids of the central sierra nevada batholith, california: nature,
  origin, and relations with the hosts}, {\emph{Lithos} {\bf 80} (2005)
  155--177}.

\bibitem{paterson2004significance}
S.~R. Paterson, G.~S. Pignotta and R.~H. Vernon, \emph{The significance of
  microgranitoid enclave shapes and orientations}, {\emph{Journal of Structural
  Geology} {\bf 26} (2004) 1465--1481}.

\bibitem{paterson2003quadruple}
S.~R. Paterson, J.~Onezime, L.~Teruya and J.~{\v{Z}}{\'a}k,
  \emph{Quadruple-pronged enclaves: their significance for the interpretation
  of multiple magmatic fabrics in plutons}, {\emph{Journal of the Virtual
  Explorer} {\bf 10} (2003) 15--30}.

\bibitem{taylor1932viscosity}
G.~I. Taylor, \emph{The viscosity of a fluid containing small drops of another
  fluid}, {\emph{Proceedings of the Royal Society of London. Series A,
  Containing papers of a mathematical and physical character} {\bf 138} (1932)
  41--48}.

\bibitem{taylor1934formation}
G.~I. Taylor, \emph{The formation of emulsions in definable fields of flow},
  {\emph{Proceedings of the Royal Society of London. Series A, containing
  papers of a mathematical and physical character} {\bf 146} (1934) 501--523}.

\bibitem{rallison1984deformation}
J.~Rallison, \emph{The deformation of small viscous drops and bubbles in shear
  flows}, {\emph{Annual review of fluid mechanics} {\bf 16} (1984) 45--66}.

\bibitem{stone1994dynamics}
H.~A. Stone, \emph{Dynamics of drop deformation and breakup in viscous fluids},
  {\emph{Annual review of fluid mechanics} {\bf 26} (1994) 65--102}.

\bibitem{guido2011shear}
S.~Guido, \emph{Shear-induced droplet deformation: Effects of confined geometry
  and viscoelasticity}, {\emph{Current opinion in colloid \& interface science}
  {\bf 16} (2011) 61--70}.

\bibitem{vananroye2007effect}
A.~Vananroye, P.~Van~Puyvelde and P.~Moldenaers, \emph{Effect of confinement on
  the steady-state behavior of single droplets during shear flow},
  {\emph{Journal of rheology} {\bf 51} (2007) 139--153}.

\bibitem{hua2013effect}
H.~Hua, Y.~Li, J.~Shin, H.-k. Song and J.~Kim, \emph{Effect of confinement on
  droplet deformation in shear flow}, {\emph{International Journal of
  Computational Fluid Dynamics} {\bf 27} (2013) 317--331}.

\bibitem{dutta2014ascent}
U.~Dutta, S.~Sarkar, A.~Baruah and N.~Mandal, \emph{Ascent modes of jets and
  plumes in a stationary fluid of contrasting viscosity}, {\emph{International
  Journal of Multiphase Flow} {\bf 63} (2014) 1--10}.

\bibitem{komrakova2014lattice}
A.~Komrakova, O.~Shardt, D.~Eskin and J.~Derksen, \emph{Lattice boltzmann
  simulations of drop deformation and breakup in shear flow},
  {\emph{International Journal of Multiphase Flow} {\bf 59} (2014) 24--43}.

\bibitem{dutta2016role}
U.~Dutta, A.~Baruah and N.~Mandal, \emph{Role of source-layer tilts in the
  axi-asymmetric growth of diapirs triggered by a rayleigh--taylor
  instability}, {\emph{Geophysical Journal International} {\bf 206} (2016)
  1814--1830}.

\bibitem{amani2019dns}
A.~Amani, N.~Balcazar, E.~Gutierrez and A.~Oliva, \emph{Dns of unequal size
  droplets collision using a moving-mesh/level-set method},  in \emph{ERCOFTAC
  workshop direct and large eddy simulation}, pp.~103--109, Springer, 2019.

\bibitem{wang2023dynamics}
K.~Wang, H.~Liang, C.~Zhao and X.~Bian, \emph{Dynamics of a droplet in shear
  flow by smoothed particle hydrodynamics}, {\emph{Frontiers in Physics} {\bf
  11} (2023) 1286217}.

\bibitem{dutta2024solitary}
U.~Dutta and N.~Mandal, \emph{Solitary waves forming pulsating thermal plumes
  and their implications for multiple eruption events in large igneous
  provinces}, {\emph{Episodes Journal of International Geoscience} {\bf 47}
  (2024) 227--239}.

\bibitem{stone1989relaxation}
H.~A. Stone and L.~G. Leal, \emph{Relaxation and breakup of an initially
  extended drop in an otherwise quiescent fluid}, {\emph{Journal of Fluid
  Mechanics} {\bf 198} (1989) 399--427}.

\bibitem{dimitrakopoulos2007gravitational}
P.~Dimitrakopoulos, \emph{Gravitational effects on the deformation of a droplet
  adhering to a horizontal solid surface in shear flow}, {\emph{Physics of
  Fluids} {\bf 19} (2007) }.

\bibitem{petford2003rheology}
N.~Petford, \emph{Rheology of granitic magmas during ascent and emplacement},
  {\emph{Annual Review of Earth and Planetary Sciences} {\bf 31} (2003)
  399--427}.

\bibitem{dingwell1996volcanic}
D.~B. Dingwell, \emph{Volcanic dilemma: flow or blow?}, {\emph{Science} {\bf
  273} (1996) 1054--1056}.

\bibitem{giordano2008viscosity}
D.~Giordano, J.~K. Russell and D.~B. Dingwell, \emph{Viscosity of magmatic
  liquids: a model}, {\emph{Earth and Planetary Science Letters} {\bf 271}
  (2008) 123--134}.

\bibitem{edwards2020magma}
M.~J. Edwards, L.~Pioli, A.~J. Harris, L.~Gurioli and S.~Thivet, \emph{Magma
  fragmentation and particle size distributions in low intensity mafic
  explosions: the july/august 2015 piton de la fournaise eruption},
  {\emph{Scientific reports} {\bf 10} (2020) 13953}.

\bibitem{toramaru2023new}
A.~Toramaru and T.~Kichise, \emph{A new model of crystallization in magmas:
  Impact of pre-exponential factor of crystal nucleation rate on cooling rate
  exponent and log-linear crystal size distribution}, {\emph{Journal of
  Geophysical Research: Solid Earth} {\bf 128} (2023) e2023JB026481}.

\bibitem{glazner2014magmatic}
A.~F. Glazner, \emph{Magmatic life at low reynolds number}, {\emph{Geology}
  {\bf 42} (2014) 935--938}.

\bibitem{williams1994microgranitic}
Q.~Williams and O.~Tobisch, \emph{Microgranitic enclave shapes and magmatic
  strain histories: constraints from drop deformation theory}, {\emph{Journal
  of Geophysical Research: Solid Earth} {\bf 99} (1994) 24359--24368}.

\bibitem{rust2003determining}
A.~Rust, M.~Manga and K.~V. Cashman, \emph{Determining flow type, shear rate
  and shear stress in magmas from bubble shapes and orientations},
  {\emph{Journal of Volcanology and Geothermal Research} {\bf 122} (2003)
  111--132}.

\bibitem{piazolo2002experimental}
S.~Piazolo and C.~W. Passchier, \emph{Experimental modeling of viscous
  inclusions in a circular high-strain shear rig: Implications for the
  interpretation of shape fabrics and deformed enclaves}, {\emph{Journal of
  Geophysical Research: Solid Earth} {\bf 107} (2002) ETG--11}.

\bibitem{hodge2012experimental}
K.~F. Hodge, G.~Carazzo and A.~M. Jellinek, \emph{Experimental constraints on
  the deformation and breakup of injected magma}, {\emph{Earth and Planetary
  Science Letters} {\bf 325} (2012) 52--62}.

\bibitem{hodge2012magmatic}
K.~F. Hodge, G.~Carazzo, X.~Montague and A.~M. Jellinek, \emph{Magmatic
  structures in the tuolumne intrusive suite, california: a new model for the
  formation and deformation of ladder dikes}, {\emph{Contributions to
  Mineralogy and Petrology} {\bf 164} (2012) 587--600}.

\bibitem{burgisser2020numerical}
A.~Burgisser, A.~Carrara and C.~Annen, \emph{Numerical simulations of magmatic
  enclave deformation}, {\emph{Journal of Volcanology and Geothermal Research}
  {\bf 392} (2020) 106790}.

\bibitem{veksler2010interfacial}
I.~V. Veksler, J.~Kahn, G.~Franz and D.~B. Dingwell, \emph{Interfacial tension
  between immiscible liquids in the system k2o-feo-fe2o3-al2o3-sio2 and
  implications for the kinetics of silicate melt unmixing}, {\emph{American
  Mineralogist} {\bf 95} (2010) 1679--1685}.

\bibitem{brackbill1992}
J.~Brackbill, D.~Kothe and C.~Zemach, \emph{A continuum method for modeling
  surface tension},
  \href{http://dx.doi.org/https://doi.org/10.1016/0021-9991(92)90240-Y}{\emph{Journal
  of Computational Physics} {\bf 100} (1992) 335--354}.

\bibitem{olsson2005}
E.~Olsson and G.~Kreiss, \emph{A conservative level set method for two phase
  flow},
  \href{http://dx.doi.org/https://doi.org/10.1016/j.jcp.2005.04.007}{\emph{Journal
  of Computational Physics} {\bf 210} (2005) 225--246}.

\bibitem{olsson2007}
E.~Olsson, G.~Kreiss and S.~Zahedi, \emph{A conservative level set method for
  two phase flow ii},
  \href{http://dx.doi.org/https://doi.org/10.1016/j.jcp.2006.12.027}{\emph{Journal
  of Computational Physics} {\bf 225} (2007) 785--807}.

\bibitem{scaillet2000phase}
B.~Scaillet, A.~Whittington, C.~Martel, M.~Pichavant and F.~Holtz, \emph{Phase
  equilibrium constraints on the viscosity of silicic magmas ii: implications
  for mafic--silicic mixing processes}, {\emph{Earth and Environmental Science
  Transactions of the Royal Society of Edinburgh} {\bf 91} (2000) 61--72}.

\end{thebibliography}\endgroup

\end{document}